\documentclass[12pt, draftclsnofoot, onecolumn]{IEEEtran}
\usepackage{epsfig,graphicx,subfigure,psfrag,amsmath,cases,bm}
\usepackage{latexsym,amssymb,amsmath,epsfig,subfigure,algorithm,mathtools}
\usepackage{algorithmic}
\usepackage{color}
\usepackage{url}
\usepackage{scrtime}

\title{Optimal 3D-Trajectory Design and Resource \\ \vspace*{-3mm}
Allocation for Solar-Powered UAV \\  \vspace*{-3mm}
Communication Systems\vspace*{-3mm}}

\author{\IEEEauthorblockN {Yan Sun,  Dongfang Xu, Derrick Wing Kwan Ng,\\
Linglong Dai, and Robert Schober\thanks{Yan Sun, Dongfang Xu, and Robert Schober are with the Institute for Digital Communications, Friedrich-Alexander-University Erlangen-N\"urnberg (FAU), Germany (email:\{yan.sun, dongfang.xu, robert.schober\}@fau.de). Derrick Wing Kwan Ng is with the School of Electrical Engineering and Telecommunications, the University of New South Wales, Australia (email: w.k.ng@unsw.edu.au). Linglong Dai is with the Department of Electronic Engineering, Tsinghua University, Beijing, China (email: daill@tsinghua.edu.cn).
This paper was presented in part at IEEE SPAWC 2018 \cite{sun2018solarUAV}.
}}\vspace*{-3mm}
}

\newtheorem{Thm}{Theorem}

\newtheorem{T-Prob}{Transformed Problem}

\DeclareMathOperator{\maxo}{maximize}
\DeclareMathOperator{\mino}{minimize}

 \newcommand{\qed}{\hfill \ensuremath{\blacksquare}}
\newtheorem{Remark}{Remark}

\newcommand{\abs}[1]{\lvert#1\rvert}
\newcommand{\norm}[1]{\lVert#1\rVert}
\begin{document}
\maketitle \vspace*{-17mm}

\begin{abstract}\vspace*{-2mm}
In this paper, we investigate the resource allocation algorithm design for multicarrier solar-powered unmanned aerial vehicle (UAV) communication systems.
In particular, the UAV is powered by solar energy enabling sustainable communication services to multiple ground users.
We study the joint design of the three-dimensional (3D) aerial trajectory and the wireless resource allocation for maximization of the system sum throughput over a given time period.
As a performance benchmark, we first consider an offline resource allocation design assuming non-causal knowledge of the channel gains. The algorithm design is formulated as a mixed-integer non-convex optimization problem taking into account the aerodynamic power consumption, solar energy harvesting, a finite energy storage capacity, and the quality-of-service (QoS) requirements of the users. Despite the non-convexity of the optimization problem, we solve it optimally by applying monotonic optimization to obtain the optimal 3D-trajectory and the optimal power and subcarrier allocation policy. Subsequently, we focus on online algorithm design which only requires real-time and statistical knowledge of the channel gains. The optimal online resource allocation algorithm is motivated by the offline scheme and entails a high computational complexity. Hence, we also propose a low-complexity iterative suboptimal online scheme based on successive convex approximation.
Our simulation results reveal that both proposed online schemes closely approach the performance of the benchmark offline scheme and substantially outperform two baseline schemes. Furthermore, our results unveil the tradeoff between solar energy harvesting and power-efficient communication. In particular, the solar-powered UAV first climbs up to a high altitude to harvest a sufficient amount of solar energy and then descents again to a lower altitude to reduce the path loss of the communication links to the users it serves.

\end{abstract}\vspace*{-4mm}
%\begin{keywords}\vspace*{-3mm}3D-Trajectory planning, solar energy harvesting, unmanned aerial vehicle systems, multicarrier systems, non-convex optimization
%\end{keywords}

\vspace*{-3mm}
\section{Introduction}
Future wireless communication systems are envisioned to provide ubiquitous and sustainable high data-rate communication services \cite{Andrews5GSurvey,book:Key5GWong}.
However, in some practical scenarios, deploying conventional terrestrial infrastructure is not cost-effective or not feasible. For example, deploying fixed base stations (BSs) in a timely and economical manner in temporary hotspots, disaster areas, and complex terrains can be challenging.
To handle this issue, aerial communication systems based on unmanned aerial vehicles (UAVs) have been proposed as a promising new paradigm to facilitate fast and highly flexible deployment of communication infrastructure due to their high maneuverability, e.g. \cite{wu2018fundamental}\nocite{Zeng16Throughput,wu2017joint,zhang2017securing,Yaliniz163DMaxUser}--\cite{Mozaffari16MaxCoverage}.
In particular, UAVs equipped with on-board wireless transceivers can fly over the target area and provide communication services.
Moreover, since UAVs enjoy high mobility, they can adapt their aerial position according to the real-time locations of the users which introduces additional spatial degrees of freedom for improving system performance.
In \cite{Zeng16Throughput}, the authors investigated UAV trajectory design for minimization of the mission completion time in multicast systems.
The authors of \cite{wu2017joint} proposed a suboptimal joint trajectory, power allocation, and user scheduling algorithm for maximization of the minimum user throughput in multi-UAV systems.
In \cite{zhang2017securing}, a suboptimal joint trajectory and power allocation algorithm was proposed for maximization of the system secrecy rate in a UAV communication system.
The placement of UAVs in the three-dimensional (3D) space for maximization of the number of served users and the coverage area was studied in \cite{Yaliniz163DMaxUser} and \cite{Mozaffari16MaxCoverage}, respectively.
However, the UAV-based communication systems considered in \cite{Zeng16Throughput}\nocite{wu2017joint,zhang2017securing,Yaliniz163DMaxUser}--\cite{Mozaffari16MaxCoverage} were powered by on-board batteries with limited energy storage capacity, leading to a constrained operation time. In fact, the UAVs in
\cite{Zeng16Throughput}\nocite{wu2017joint,zhang2017securing,Yaliniz163DMaxUser}--\cite{Mozaffari16MaxCoverage} are required to return to their home base frequently for recharging their batteries.
Hence, these designs cannot guarantee stable and sustainable communication services which may create a system performance bottleneck.

To overcome these shortcomings, solar-powered UAVs have received significant attention due to their potential to realize perpetual flight \cite{oettershagen2016perpetual,morton2015solar}.
In particular, solar panels equipped at the UAVs can harvest solar energy and convert it to electrical energy enabling long endurance flights.
For instance, the authors of \cite{oettershagen2016perpetual} and \cite{morton2015solar} have developed solar-powered UAV prototypes and demonstrated the possibility of continuous flight for $28$ hours.
However, the amount of harvested solar energy depends on the flight altitude of the UAV.
In particular, the intensity of solar energy significantly decreases if the light passes through clouds resulting in a reduced received solar energy flux at the solar panel \cite{duffie2013solar,kokhanovsky2004optical}.
Thus, UAVs flying above clouds can generally harvest more solar energy than those flying below clouds.
In \cite{Lee2017PathSolarUAV}, the authors studied the optimal trajectory of solar-powered UAVs for maximization of the harvested solar power.
However, \cite{Lee2017PathSolarUAV} focused only on the flight control of solar-powered UAVs.
The proposed design did not consider the influence of clouds on energy harvesting based communications.
Hence, a higher flight altitude was always preferable as more energy could be harvested.
However, since higher flight altitudes lead to a more severe path loss for air-to-ground communications, there is a non-trivial tradeoff between harvesting more solar energy and improving communication performance.
This tradeoff does not exist in conventional UAV communication systems and the results derived in \cite{Zeng16Throughput}\nocite{wu2017joint,zhang2017securing,Yaliniz163DMaxUser}--\cite{Mozaffari16MaxCoverage} are only applicable to non-energy harvesting UAV communication systems.
In our previous work \cite{sun2018solarUAV}, we studied the resource allocation for multicarrier (MC) solar-powered UAV communication systems. In particular, a suboptimal algorithm for joint 3D positioning of the UAV, power adaption, and subcarrier allocation for maximization of the system sum throughput was proposed.
However, the constant aerodynamic power consumption model adopted in \cite{sun2018solarUAV} is only valid when the flight velocity is constant. In practice, the aerodynamic power consumption depends on the flight velocity and contributes significantly to the overall power consumption of the UAV. Hence, assuming constant aerodynamic power consumption is not valid for realistic UAV systems with non-constant speed.
Besides, the resource allocation design in \cite{sun2018solarUAV} focused on the positioning of the UAV and the resulting algorithm cannot be applied for optimization of 3D aerial trajectory.
Furthermore, in UAV-based communication systems, satisfying the quality-of-service (QoS) requirements of the users is of paramount importance \cite{wu2018fundamental}. However, in \cite{sun2018solarUAV}, the QoS requirements of the users were not taken into account for resource allocation design.
In practice, the coupling between trajectory optimization, aerodynamic power control, and QoS guarantees for communication complicates the optimal resource allocation design for solar-powered MC-UAV communication systems.
Moreover, most of the existing trajectory and resource allocation designs for UAV-based communication systems are suboptimal \cite{sun2018solarUAV}, \cite{Zeng16Throughput}\nocite{wu2017joint,zhang2017securing,Yaliniz163DMaxUser}--\cite{Mozaffari16MaxCoverage}, and the performance gap between these designs and the optimal one is still unknown. In fact, the optimal joint trajectory and resource allocation algorithm design for solar-powered MC-UAV communication systems with QoS constraints is still an open problem.

In this paper, we address the above issues. To this end, we first focus on the offline case where non-causal knowledge of the channel gains is available. The joint trajectory and resource allocation algorithm design for solar-powered MC-UAV communication systems is formulated as a combinatorial non-convex optimization problem for maximization of the system sum throughput over a finite horizon. Our problem formulation takes into account the solar energy harvesting, the aerodynamic power consumption, the dynamics of the on-board energy storage, and the QoS requirements of the users. Although the considered problem is non-convex and difficult to tackle, we solve it optimally by exploiting the theory of monotonic optimization \cite{tuy2000monotonic,zhang2013monotonic} and obtain the jointly optimal trajectory and power and subcarrier allocation policy. Besides, we also investigate the online resource allocation algorithm design which requires only causal knowledge of the channel states. The structure of the derived offline solution serves as a building block for the design of the optimal online resource allocation algorithm. Since the optimal online policy entails a high computational complexity,
we also develop a low-complexity suboptimal online algorithm based on successive convex optimization which is shown to achieve a close-to-optimal performance.
Simulation results reveal that the performance of the two proposed online resource allocation schemes closely approaches that of the offline scheme. Besides, our simulation results show that the proposed solar-powered MC-UAV systems achieve a significant improvement in average system throughput compared to two baseline schemes.

\vspace*{-5mm}
\section{Notation and System Model}
In this section, we present the considered MC-UAV communication system model as well as the adopted solar energy harvesting and UAV aerodynamic power consumption models. However, first we introduce some notation.

\vspace*{-3mm}
\subsection{Notation}%
We use boldface lower case letters to denote vectors.
$\mathbb{C}$ denotes the set of complex numbers; $\mathbb{R}^{N\times 1}$ denotes the set of all $N\times 1$ vectors with real entries; $\mathbb{R}^+$ denotes the set of non-negative real numbers; $\mathbb{Z}^{N\times 1}$ denotes the set of all $N\times 1$ vectors with integer entries; $\abs{\cdot}$ and $\norm{\cdot}$ denote the absolute value of a complex scalar and the Euclidean vector norm, respectively;
${\cal E}\{\cdot\}$ denotes statistical expectation;  ${\mathrm{Var}}\{\cdot\}$ denotes the statistical variance; the circularly symmetric complex Gaussian distribution with mean $\mu$ and variance $\sigma^2$ is denoted by ${\cal CN}(\mu,\sigma^2)$; and $\sim$ stands for ``distributed as"; $\nabla_{\mathbf{x}} f(\mathbf{x})$ denotes the gradient vector of function $f(\mathbf{x})$, i.e., its components are the partial derivatives of $f(\mathbf{x})$.

\vspace*{-3mm}
\subsection{MC-UAV Communication System Model}%
\begin{figure}
\centering\vspace*{-5mm}
\includegraphics[width=3.3in]{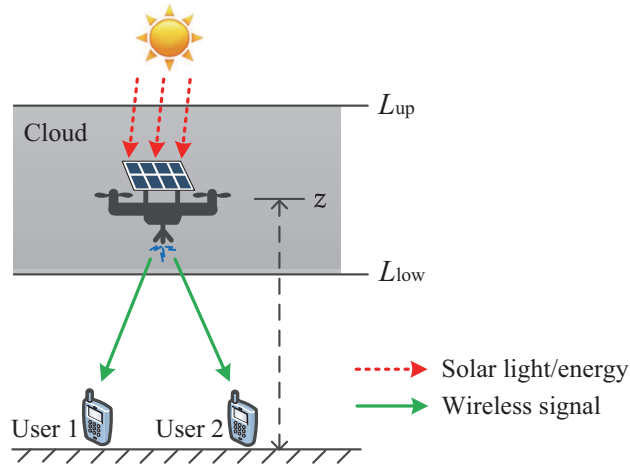}\vspace*{-3mm}
\caption{A solar-powered MC-UAV communication system with one UAV transmitter and $K=2$ downlink users.}
\label{fig:system_model}\vspace*{-5mm}
\end{figure}

The considered MC-UAV wireless communication system comprises one rotary-wing UAV-mounted transmitter \cite{Tech:LTEUAVQualcomm} and $K$ downlink users. The UAV-mounted transmitter and the downlink users are single-antenna half-duplex devices, cf. Figure \ref{fig:system_model}.
The UAV is equipped with solar panels which harvest solar energy and convert it to electrical energy. The harvested energy is stored in the on-board battery and is used for providing communication services and powering the flight operation of the UAV.
We focus on a wideband system where the system bandwidth $\mathcal{W}$ Hz is divided into $N_{\mathrm{F}}$ orthogonal subcarriers. We assume that each subcarrier can be allocated to at most one user\footnote{The considered system can be extended to the case of non-orthogonal multiple access (NOMA), where multiple users are multiplexed on each subcarrier, by applying the power and subcarrier allocation design framework proposed in \cite{sun2016optimalJournal}.}.

To facilitate the trajectory planning of the UAV, we employ the discrete path planning approach \cite{enright1992discrete,betts1998survey}.
In particular, the trajectory of the UAV during the operation time period $T$ is discretized into $N_{\mathrm{T}}$ waypoints and the period $T$ is divided into $N_{\mathrm{T}}$ equal-length time slots. The duration of each time slot is $\Delta_{\mathrm{T}}$ such that $T=N_{\mathrm{T}}\Delta_{\mathrm{T}}$. Note that the location of the UAV can be assumed to be approximately unchanged during each time slot when $\Delta_{\mathrm{T}}$ is chosen sufficiently small, cf. \cite{enright1992discrete,betts1998survey}.
%%%%%%%%%%%
In a given time slot $n\in\{1,\ldots,N_{\mathrm{T}}\}$, the path loss of the communication link between the UAV and user $k \in \{1,\ldots,K\}$ is modeled as $\zeta\norm{\mathbf{r}[n]-\mathbf{r}_k}^{-2}$,
where $\mathbf{r}[n]=(x[n],y[n],z[n])$ and $\mathbf{r}_k=(x_k,y_k,0)$ specify the 3D Cartesian coordinates of the UAV in time slot $n$ and user $k$, respectively. In particular, $(x[n],y[n])$ and $(x_k,y_k)$ are the horizontal coordinates of the UAV and user $k$, respectively, and $z[n]$ denotes the altitude of the UAV. Besides, $\zeta=(\frac{c}{4\pi f_{\mathrm{0}}})^2$, where $c$ is the speed of light and $f_{\mathrm{0}}$ is the center frequency of the carrier signal.

Therefore, in a given scheduling time slot $n$, the received signal at downlink user $k$ on subcarrier $i\in\{1,\ldots, N_{\mathrm{F}}\}$ is given by \vspace*{-2mm}
\begin{eqnarray}
u_k^i [n]= \frac{\sqrt{\zeta p_k^i[n]}h_k^i[n]}{\norm{\mathbf{r}[n]-\mathbf{r}_k}} d_k^i[n] + n_k^i[n],
\end{eqnarray}
where $d_k^i[n]\in\mathbb{C}$ denotes the data symbol transmitted from the UAV to user $k$ on subcarrier $i$ in time slot $n$ and we assume ${\cal E}\{\abs{d_k^i[n]}^2\}=1$ without loss of generality.
$p_k^i[n]\in\mathbb{R}^+$ and $h_k^i[n]\in\mathbb{C}$ denote the transmit power and the channel gain\footnote{The coefficient $h_k^i[n]$ can be assumed to be unchanged during each time slot if the displacement of the UAV during a time slot is sufficiently small. For example, for $\mathcal{W}=5$ $\mathrm{MHz}$ total bandwidth and center frequency $700$ $\mathrm{MHz}$, $h_k^i[n]$ can be assumed to be unchanged in time slot $n$ if the  displacement of the UAV in time slot $n$ is smaller than half a wavelength of the carrier signal, i.e., $\frac{c}{2 f_{\mathrm{0}}}=0.214$ m \cite{stutzman2012antenna}.}  from the UAV to user $k$ on subcarrier $i$ in time slot $n$, respectively.
In particular, channel coefficient $h_k^i[n]$ captures the shadowing and the small-scale fading effects due to multipath propagation \cite{Hourani14Model,khuwaja2018survey}. In fact, according to field measurements, reflection and scattering of the UAV transmit signal occurs in air-to-ground communication links \cite{Hourani14Model,khuwaja2018survey} although this is often neglected in the literature \cite{Zeng16Throughput}\nocite{wu2017joint}--\cite{zhang2017securing}, \cite{Xu2018UAVPowerTransfer}\nocite{zeng2018energy}--\cite{wu2018common}.
$n_k^i[n]\sim{\cal CN}(0,\mathcal{N}_0 \mathcal{B})$ denotes the complex additive white Gaussian noise (AWGN) on subcarrier $i$ at user $k$, where $\mathcal{N}_0$ denotes the noise power spectral density and $\mathcal{B}=\frac{\mathcal{W}}{N_{\mathrm{F}}}$ is the bandwidth of one subcarrier.
We note that clouds have negligible impact on radio frequency (RF) signals for carrier frequencies $f_{\mathrm{0}}$ below $10$ GHz \cite{ITUAttenuCloud}.
\begin{Remark}
The considered UAV-mounted transmitter can serve as an aerial mobile BS providing stable and sustainable communication services to multiple ground users in temporary hotspots, disaster areas, and complex terrains \cite{Zeng16UAVMag}, etc. The backhaul between the UAV and the core network can be established e.g. via out-of-band free space optical (FSO) links \cite{alzenad2018fso,najafi2017statistical}.
\end{Remark}

\vspace*{-3mm}
\subsection{Solar Energy Harvesting}
\begin{figure}
\centering\vspace*{-5mm}
\includegraphics[width=3.4in]{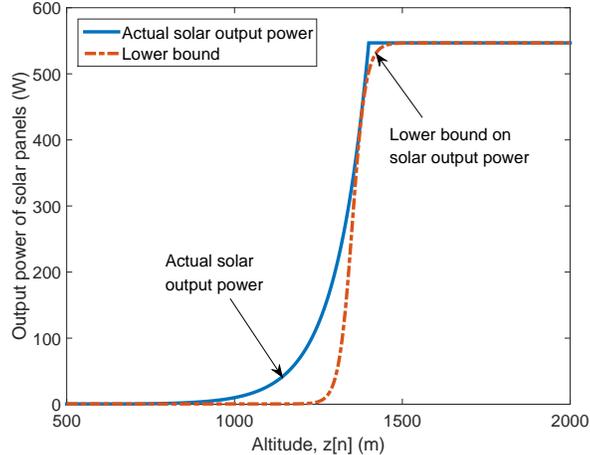}\vspace*{-3mm}
\caption{Illustration of the actual solar output power $P^{\mathrm{solar}}\big(z[n]\big)$ and the corresponding lower bound $\underline{P}^{\mathrm{solar}}\big(z[n]\big)$.}
\label{fig: solar_approx_curve}\vspace*{-5mm}
\end{figure}

The considered MC-UAV communication system is powered by the harvested solar energy. In general, the amount of harvested solar energy is affected by clouds \cite{duffie2013solar,kokhanovsky2004optical}.
In particular, the harvested solar energy is reduced if there is a cloud between the sun and the solar panel. The attenuation of the solar light passing through a cloud can be modeled as \cite{kokhanovsky2004optical}: \vspace*{-1mm}
\begin{equation}
\varphi(d^{\mathrm{cloud}})=e^{-\beta_c d^{\mathrm{cloud}}},
\end{equation}
where $\beta_c \ge 0$ denotes the absorption coefficient modeling the optical characteristics of the cloud and $d^{\mathrm{cloud}}$ denotes the distance that the solar light propagates through the cloud. Therefore, the electrical output power of a solar panel at altitude $z$ is modeled by the following function \cite{duffie2013solar}\nocite{kokhanovsky2004optical}--\cite{Lee2017PathSolarUAV}: \vspace*{-2mm}
\begin{equation}\label{solar_power}
\hspace*{-2mm}P^{\mathrm{solar}}\big(z[n]\big)\hspace*{-0.5mm}=
\hspace*{-0.5mm} \left\{
\begin{array}{lcl}
\hspace*{-2mm} \eta S G, & &\hspace*{-3mm} z[n] \hspace*{-0.5mm} \ge \hspace*{-0.5mm} L_{\mathrm{up}},\\[-0mm]
%%%%
\hspace*{-2mm} \eta S G e^{-\beta_c (L_{\mathrm{up}}-z[n])}, & &\hspace*{-3mm} L_{\mathrm{low}} \hspace*{-0.5mm} \le \hspace*{-0.5mm} z[n] \hspace*{-0.5mm} < \hspace*{-0.5mm} L_{\mathrm{up}},\\[-0mm]
%%%%%%
\hspace*{-2mm} \eta S G e^{-\beta_c (L_{\mathrm{up}}-L_{\mathrm{low}})}, & &\hspace*{-3mm} z[n] \hspace*{-0.5mm} < \hspace*{-0.5mm} L_{\mathrm{low}},
\end{array}\right.
\end{equation}
where $\eta$ and $S$ are constants representing the energy harvesting efficiency and the equivalent area of the solar panels, respectively. Constant $G$ denotes the average solar radiation intensity on earth. $L_{\mathrm{up}}$ and $L_{\mathrm{low}}$ are the altitudes of the upper and lower boundaries of the cloud, respectively, cf. Figure \ref{fig:system_model}.
%%%%%
$P^{\mathrm{solar}}\big(z[n]\big)$ is a piecewise function where the output power of the solar panels increases exponentially with the altitude in the cloud and becomes a constant when the solar panels are below or above the cloud.
The non-smoothness of the solar output power significantly complicates resource allocation algorithm design.
%%%%%%
However, it can be shown that the piecewise function $P^{\mathrm{solar}}\big(z[n]\big)$ is always bounded below by
\begin{equation}\label{solar_bound}
\underline{P}^{\mathrm{solar}}\big(z[n]\big) = \frac{C_1}{1+e^{-k_{c}(z[n]-\alpha)}}+ C_2,
\end{equation}
where $C_1 = \eta S G(1-e^{-\beta_c (L_{\mathrm{up}}-L_{\mathrm{low}})})$, $C_2 = \eta S G e^{-\beta_c (L_{\mathrm{up}}-L_{\mathrm{low}})}$, and $k_{c} \ge 0$ and $\alpha \ge 0$ are parameters to adjust the gap between the lower bound on the solar output power $\underline{P}^{\mathrm{solar}}\big(z[n]\big)$ and the actual solar output power $P^{\mathrm{solar}}\big(z[n]\big)$.
The actual solar output power and the lower bound are illustrated in Figure \ref{fig: solar_approx_curve}, where the adopted system parameters are specified in Table \ref{tab:parameters}. As the actual solar output power is a piecewise non-smooth function, we adopt the lower bound on the solar output power in the sequel to facilitate resource allocation design for solar-powered UAV communication systems.

For the considered solar-powered UAV communication system, we note that there is a fundamental tradeoff between  solar energy harvesting and improving communication performance. In particular, the UAV can harvest more solar energy by climbing up to higher altitudes. However, flying at a higher altitude leads to a larger path loss for the communication links between the UAV and the users which causes a degradation of the system performance.

\vspace*{-3mm}
\subsection{Aerodynamic Power Consumption}
In each time slot, we assume that the UAV is in a quasi-static equilibrium condition \cite{enright1992discrete,zeng2018energy}. This means that the UAV moves smoothly with a small acceleration and the cruising speed is assumed to be a constant during each time slot.
In particular, we define $\mathbf{v}[n]=\big(v_{x}[n],v_{y}[n],v_{z}[n]\big)$ as the velocity of the UAV during the $n$-th time slot, where $v_{x}[n]$, $v_{y}[n]$, and $v_{z}[n]$ are the velocity components of $\mathbf{v}[n]$ in 3D Cartesian coordinates and are constant during time slot $n$.
According to the classical aircraft dynamics of rotary-wing UAVs, the aerodynamic power consumption of a UAV can be modeled as a linear sum of the induced power\footnote{The induced power is the minimum required power to maintain the UAV levitating in the air \cite{seddon2011basic,Bangura17Thrust}.} for level flight, the power for vertical flight, and the profile power related the blade drag \cite{seddon2011basic,Bangura17Thrust}.
In particular, the induced power for level flight in time slot $n$ is modeled as \cite[Eq. (7.10)]{seddon2011basic}: \vspace*{-3mm}
\begin{eqnarray}\label{Plevel}
P_{\mathrm{level}}[n] = \frac{W^2}{\sqrt{2}\rho A} \cdot \frac{1}{\sqrt{\norm{(v_x[n], v_y[n])}^2 + \sqrt{\norm{(v_x[n], v_y[n])}^4 + 4 V_{\mathrm{h}}^4}}},
\end{eqnarray}
where $W=mg$ is the weight of the UAV and $m$ and $g$ denote the mass of the UAV and the gravitational acceleration, respectively. $\rho$ is the density of air and $A$ is the total area of the UAV rotor disks \cite{hoffmann07quadrotorTheroyExp}. $\norm{(v_x[n], v_y[n])}$ represents the horizontal speed of the UAV and constant $V_{\mathrm{h}}=\sqrt{\frac{W}{2\rho A}}$ parameterizes the required power for hovering\footnote{The required power for hovering is $P_{\mathrm{hover}}=\frac{W^2}{\sqrt{2}\rho A} \cdot \frac{1}{\sqrt{ \sqrt{4 V_{\mathrm{h}}^4}}} = \frac{W^2}{\sqrt{2}\rho A} \cdot \frac{1}{\sqrt{2}} \cdot \sqrt{\frac{2\rho A}{W}}= \frac{W^{3/2}}{\sqrt{2\rho A}}$.} \cite{seddon2011basic,hoffmann07quadrotorTheroyExp}. Eq. \eqref{Plevel} implies that less power is consumed during level flight compared to hovering.
%%%%%%%%%%%%%%%%%%%%%%%%%%%%%%%%%%%%%%%%%%%%%%%%%%%%%%%%%%%%%
Besides, the power consumption for vertical flight in time slot $n$ is modeled as \cite[Eq. (7.12)]{seddon2011basic}:\vspace*{-2mm}
\begin{eqnarray}\label{Pclimb}
P_{\mathrm{vertical}}[n] = W v_{z}[n].
\end{eqnarray}
From \eqref{Pclimb}, we note that climbing flight consumes more power than hovering and descending flight\footnote{For descending flights, $P_{\mathrm{vertical}}[n]$ in \eqref{Pclimb} is negative as gravity leads to power savings.}.
%%%%%%%%%%%%%%%%%%%%%%%%%%
In addition, the blade drag profile power in time slot $n$ is modeled as \cite[Eq. (7.1)]{seddon2011basic}: \vspace*{-0mm}
%\begin{eqnarray}\label{Pdrag}
%P_{\mathrm{drag}}[n] = \frac{1}{8} C_{\mathrm{D0}} \rho A \Big(\frac{\norm{(v_x[n], v_y[n])}}{\mu_{\mathrm{b}}}\Big)^3,
%\end{eqnarray}
\begin{eqnarray}\label{Pdrag}
P_{\mathrm{drag}}[n] = \frac{1}{8} C_{\mathrm{D0}} \rho A \norm{(v_x[n], v_y[n])}^3,
\end{eqnarray}
where $C_{\mathrm{D0}}$ is the profile drag coefficient which depends on the geometry of the rotor blades. We note that the drag profile power is proportional to the horizontal velocity and independent of the vertical velocity.

In summary, the aerodynamic power consumption of the UAV in time slot $n$ can be modeled as: \vspace*{-3mm}
\begin{eqnarray}\label{aero_power_consumption}
\hspace*{-4mm} P_{\mathrm{UAV}}[n] \hspace*{-2mm} &=& \hspace*{-2mm} P_{\mathrm{level}}[n] + P_{\mathrm{vertical}}[n] + P_{\mathrm{drag}}[n] \notag \\
\hspace*{-2mm} &=& \hspace*{-2mm}  \underbrace{\frac{\varrho_1}{\sqrt{\norm{(v_x[n], v_y[n])}^2 \hspace*{-0.8mm} + \hspace*{-0.8mm} \sqrt{\norm{(v_x[n], v_y[n])}^4 \hspace*{-0.8mm}+\hspace*{-0.8mm} 4 V_{\mathrm{h}}^4}}}}_{\text{level flight power consumption}}
\hspace*{-0mm} + \hspace*{-2mm} \underbrace{W v_{z}[n]}_{\substack{\text{vertical flight}\\\text{power consumption}}} \hspace*{-3mm} + \hspace*{1mm} \underbrace{\varrho_2 \norm{(v_x[n], v_y[n])}^3}_{\text{drag power consumption}},
\end{eqnarray}
where $\varrho_1 = \frac{W^2}{\sqrt{2}\rho A}$ and $\varrho_2 = \frac{1}{8} C_{\mathrm{D0}} \rho A$.

\vspace*{-3mm}
\section{Offline Trajectory and Resource Allocation Design}
\label{Offline-section}
In this section, we design the optimal trajectory and resource allocation based on an offline approach by assuming non-causal knowledge of the channel gains. After defining the adopted performance metric, we formulate the design as a non-convex optimization problem and solve it optimally using monotonic optimization.
\vspace*{-2mm}
\subsection{Achievable Data Rate}
In time slot $n$, assuming subcarrier $i$ is allocated to user $k$, the achievable data rate (bits/s) on subcarrier $i$ is given by: \vspace*{-3mm}
\begin{equation}\label{rate_k}
R_k^i[n](\mathbf{p},\mathbf{s},\mathbf{r})=s_{k}^i[n] \mathcal{B} \log_2 \Big( 1 + \frac{H_k^i[n] p_k^i[n]}{\norm{\mathbf{r}[n]-\mathbf{r}_k}^2} \Big),
\end{equation}
where $H_k^i[n]=\frac{\zeta\abs{h_k^i[n]}^2}{\mathcal{N}_0 \mathcal{B}}$. Variable $s_k^i [n]\in \{0,1\}$ is the binary subcarrier allocation indicator. Specifically, $s_k^i [n]= 1$ if user $k$ is allocated to subcarrier $i$ and $s_k^i [n] = 0$, otherwise.
$\mathbf{p}\in\mathbb{R}^{N_{\mathrm{T}}N_{\mathrm{F}}K \times 1}$, $\mathbf{s}\in\mathbb{Z}^{N_{\mathrm{T}}N_{\mathrm{F}}K \times 1}$, and $\mathbf{r}\in\mathbb{R}^{3N_{\mathrm{T}}\times 1}$ are the collections of all $p_k^i[n]$, $s_k^i[n]$, and $\mathbf{r}[n]$, respectively.

\vspace*{-2mm}
\subsection{Optimization Problem Formulation}
In this paper, we maximize the system sum throughput (bits/s/Hz) during a period of $N_{\mathrm{T}}$ time slots. The trajectory and the power and subcarrier allocation policy are obtained by solving the following optimization problem: \vspace*{-2mm}
\begin{eqnarray}
\label{prob}
&&\hspace*{-1mm} \underset{\mathbf{p},\mathbf{s},\mathbf{r},\mathbf{v},\mathbf{q}}{\maxo} \,\, \,\, \frac{1}{N_{\mathrm{F}} \mathcal{B}} \overset{N_{\mathrm{T}}}{\underset{n = 1}{\sum}} \overset{N_{\mathrm{F}}}{\underset{i = 1}{\sum}} \overset{K}{\underset{k = 1}{\sum}} s_{k}^i[n] \mathcal{B}\log_2 \Big( 1 + \frac{H_k^i[n] p_k^i[n]}{\norm{\mathbf{r}[n]-\mathbf{r}_k}^2} \Big)  \\[-2mm]
\notag\mbox{s.t.}\hspace*{1mm}
&&\hspace*{-7mm}   \mbox{C1: } \Bigg[\overset{N_{\mathrm{F}}}{\underset{i = 1}{\sum}} \overset{K}{\underset{k = 1}{\sum}} \frac{1}{\varepsilon} s_{k}^i[n] p_k^i[n] \hspace*{-0.5mm}  + \hspace*{-0.5mm} P_{\mathrm{UAV}}[n] + P_{\mathrm{static}}\Bigg]\Delta_{\mathrm{T}} \hspace*{-0.5mm} \le \hspace*{-0.5mm}  q[n], \forall n, \notag \\[-2.5mm]
%%%%
&&\hspace*{-7mm}   \mbox{C2: } q[n+1] \le q[n] + \underline{P}^{\mathrm{solar}}\big(z[n]\big)\Delta_{\mathrm{T}} - \Bigg[ \overset{N_{\mathrm{F}}}{\underset{i = 1}{\sum}} \overset{K}{\underset{k = 1}{\sum}} \frac{1}{\varepsilon} s_{k}^i[n] p_k^i[n] \hspace*{-0.5mm}  + \hspace*{-0.5mm} P_{\mathrm{UAV}}[n] + P_{\mathrm{static}}\Bigg]\Delta_{\mathrm{T}}, \forall n, \notag \\[-1mm]
%%%%
&&\hspace*{-7mm}   \mbox{C3: } \mathbf{r}[n+1] = \mathbf{r}[n] + \mathbf{v}[n+1]\Delta_{\mathrm{T}}, \, \forall n,
  \hspace*{23mm}    \mbox{C4: } \norm{\mathbf{v}[n+1] - \mathbf{v}[n]} \le a_{\mathrm{max}}\Delta_{\mathrm{T}}, \forall n,  \notag\\[-1mm]
%%%%%%%%%%%%%%%%%%%%%%%%%%%%%%%
&&\hspace*{-7mm}  \mbox{C5: } p_k^i[n] \hspace*{-0.5mm}\ge \hspace*{-0.5mm} 0, \forall i,k,n,
  \hspace*{22.5mm}   \mbox{C6: } 0 \hspace*{-0.5mm} \le \hspace*{-0.5mm} q[n] \hspace*{-0.5mm} \le \hspace*{-0.5mm}  q_{\mathrm{max}}, \forall n,
  \hspace*{3.5mm}   \mbox{C7: } q[1] \hspace*{-0.2mm} =  \hspace*{-0.2mm} q_{0},  q[\hspace*{-0.5mm}N_{\mathrm{T}}\hspace*{-0.5mm}+\hspace*{-0.5mm}1\hspace*{-0.5mm}] \hspace*{-0.2mm} \ge \hspace*{-0.2mm} q_{\mathrm{end}}, \notag \\[-1.5mm]
%%%%%%%%%%%%%%%%%%%%%%
&&\hspace*{-7mm}  \mbox{C8: } \norm{\big(v_x[n], v_y[n]\big)} \le V_{\mathrm{max}}^{\mathrm{xy}},  \forall n,  \quad
  \hspace*{0mm}   \mbox{C9: } \abs{v_z [n]} \le V_{\mathrm{max}}^{\mathrm{z}},  \forall n, \quad
  \hspace*{0.5mm}   \mbox{C10: } z_{\mathrm{min}} \le z[n] \le z_{\mathrm{max}},  \forall n, \notag \\[-1.5mm]
%%%%%%%%%%%%%%%%%%%%%%%%%%%
&&\hspace*{-7mm}  \mbox{C11: } \overset{N_{\mathrm{F}}}{\underset{i = 1}{\sum}} \overset{K}{\underset{k = 1}{\sum}}  s_{k}^i[n] p_k^i[n] \hspace*{-0.5mm} \le \hspace*{-0.5mm} P_{\mathrm{max}},
  \hspace*{7.5mm}   \mbox{C12: } \overset{K}{\underset{k=1}{\sum}} s_{k}^i[n] \hspace*{-1mm}\le\hspace*{-1mm} 1, \forall i,n,
  \hspace*{3.5mm}\mbox{C13: } s_{k}^i[n] \hspace*{-1mm}\in\hspace*{-1mm} \{0,1\}, \forall i,k,n,  \notag\\[-2.5mm]
%%%%%%%%%%%%%%%%%%%%%%%%%%%
&&\hspace*{-7mm}  \mbox{C14: } \overset{N_{\mathrm{F}}}{\underset{i = 1}{\sum}} s_{k}^i[n] \mathcal{B}\log_2 \Big( 1 + \frac{H_k^i[n] p_k^i[n]}{\norm{\mathbf{r}[n]-\mathbf{r}_k}^2} \Big) \ge R^{\mathrm{req}}_k, \forall k,n, \notag
\end{eqnarray}
Constraint C1 is the energy constraint of the UAV in each time slot where constant $0 < \varepsilon < 1$ is the efficiency of the power amplifier, $q[n]\in\mathbb{R}$ is the available energy stored in the on-board battery of the UAV in time slot $n$, and $P_{\mathrm{static}}$ denotes the static power consumed for maintaining the operation of the UAV.
Constraint C2 is imposed since the available energy of the on-board battery in time slot $n+1$ is determined by the harvested solar energy and the energy consumption in time slot $n$.
Constraints C3 and C4 restrict the maximum displacement and the change of velocity of the UAV in each time slot, respectively, where $a_{\mathrm{max}}$ denotes the maximum possible acceleration.
Constraint C5 is the non-negative transmit power constraint.
Constraint C6 restricts the maximum energy storage capacity $q_{\mathrm{max}}$ of the on-board battery. Constraint C7 specifies the available initial energy $q_{\mathrm{0}}$ and the required remaining energy $q_{\mathrm{end}}$ in the on-board battery before the first and after the last time slot, respectively. We note that imposing a constraint on $q_{\mathrm{end}}$ is necessary for providing sustainable communication service since the UAV needs sufficient energy to start the subsequent flight period and for the return flight to the home base\footnote{Returning the UAV to a home base before the end of period $T$ can also be accomplished by adding a linear constraint $\mathbf{r}[N_{\mathrm{T}}]=\mathbf{r}_0$ in problem \eqref{prob}, where $\mathbf{r}_0$ denotes the 3D coordinates of the home base.}.
$V_{\mathrm{max}}^{\mathrm{xy}}$ in constraint C8 and $V_{\mathrm{max}}^{\mathrm{z}}$ in C9 denote the maximum horizontal and vertical speeds of the UAV, respectively.
Constraint C10 restricts the minimum flight altitude $z_{\mathrm{min}}$ and the maximum flight altitude $z_{\mathrm{max}}$ of the UAV which may be imposed by government regulations.
$P_{\mathrm{max}}$ in constraint C11 denotes the maximum transmit power of the UAV-mounted transmitter to meet a desired transmit spectrum mask. Constraints C12 and C13 are imposed to guarantee that each subcarrier is allocated to at most one user. Constraint C14 imposes a minimum required constant data rate of $R^{\mathrm{req}}_k$ for user $k$ in each time slot.
Besides, $\mathbf{v}\in\mathbb{R}^{3N_{\mathrm{T}}\times 1}$ and $\mathbf{q}\in\mathbb{R}^{N_{\mathrm{T}} \times 1}$ are the collections of all $\mathbf{v}[n]=\big(v_{x}[n],v_{y}[n],v_{z}[n]\big)$ and $q[n]$, respectively.

%\begin{Prop}
%The
%\end{Prop}

Problem \eqref{prob} is a mixed-integer combinatorial non-convex optimization problem and very difficult to solve.
In particular,  the non-convex combinatorial objective function, the non-convex constraint functions in C1, C2, and C14, and the binary selection constraint C13 are obstacles for the design of an efficient offline trajectory and resource allocation algorithm. Nevertheless, despite these challenges, in the next section, we will provide the optimal solution to problem \eqref{prob}.

\vspace*{-4mm}
\subsection{Optimal Solution} \label{Offline_optimal_section}
In this section, we solve problem \eqref{prob} optimally by applying monotonic optimization theory \cite{tuy2000monotonic,zhang2013monotonic}.
To facilitate the presentation, we define $\tilde{p}_k^i[n] = s_{k}^i[n]p_k^i[n]$ and thereby the achievable data rate on subcarrier $i$ in \eqref{rate_k} can be rewritten as:
\vspace*{-2mm}
\begin{eqnarray} \label{rate-k-eqv}
\tilde{R}_k^i[n](\tilde{\mathbf{p}},\mathbf{r})&=& \mathcal{B}\log_2 \Big( 1 + \frac{\frac{H_k^i[n]}{\norm{\mathbf{r}[n]-\mathbf{r}_k}^2}s_{k}^i[n]p_k^i[n] }{\xi \sum_{j\neq k}^{K} \frac{H_k^i[n]}{\norm{\mathbf{r}[n]-\mathbf{r}_k}^2}s_{j}^i[n]p_j^i[n] + 1} \Big) \notag \\[-1mm]
&=& \mathcal{B}\log_2 \Big( 1 + \frac{H_k^i[n] \tilde{p}_k^i[n] }{\xi\sum_{j\neq k}^{K} H_k^i[n]\tilde{p}_j^i[n] + \norm{\mathbf{r}[n]\hspace*{-0.5mm}-\hspace*{-0.5mm}\mathbf{r}_k}^2} \Big),
\end{eqnarray}
where $\xi \gg 1$ is a penalty factor and $\tilde{\mathbf{p}}\in\mathbb{R}^{N_{\mathrm{T}}N_{\mathrm{F}}K \times 1}$ is the collection of all $\tilde{p}_k^i[n]$.
In particular, the term $\xi \sum_{j\neq k}^{K} H_k^i[n]\tilde{p}_j^i[n]$ represents the co-channel multiuser interference at the receiver of user $k$ if multiple users are multiplexed on subcarrier $i$ in time slot $n$. Specifically, if a given subcarrier allocation policy satisfies constraints C12 and C13, $\xi \sum_{j\neq k}^{K} H_k^i[n]\tilde{p}_j^i[n]=0$.
Hence, \eqref{rate-k-eqv} is equivalent to \eqref{rate_k} for all feasible solutions.
In other words, $\xi \sum_{j\neq k}^{K} H_k^i[n]\tilde{p}_j^i[n]$ acts as a penalty term to penalize the objective function for any violation of constraints C12 and C13.

Then, adopting the utility function in \eqref{rate-k-eqv}, we rewrite the original problem \eqref{prob} as:\vspace*{-1mm}
\begin{eqnarray}\label{equiv-prob}
&&\hspace*{-1mm} \underset{\tilde{\mathbf{p}},\mathbf{r},\mathbf{v},\mathbf{q},\bm{\theta}}{\maxo} \,\, \,\, \overset{N_{\mathrm{T}}}{\underset{n = 1}{\sum}} \overset{N_{\mathrm{F}}}{\underset{i = 1}{\sum}} \overset{K}{\underset{k = 1}{\sum}} \log_2 \Big( 1 + \frac{ H_k^i[n] \tilde{p}_k^i[n] }{\xi \sum_{j\neq k}^{K} H_k^i[n]\tilde{p}_j^i[n] + \theta_k[n]} \Big)  \\[-2mm]
\hspace*{-1mm}\notag\mbox{s.t.}\hspace*{1mm}
&&\hspace*{-7mm}  \mbox{C3, C4, C6--C10,}
\hspace*{37mm}  \mbox{C1: } \Bigg[\overset{N_{\mathrm{F}}}{\underset{i = 1}{\sum}} \overset{K}{\underset{k = 1}{\sum}} \frac{1}{\varepsilon} \tilde{p}_k^i[n] \hspace*{-0.5mm}  + \hspace*{-0.5mm} P_{\mathrm{UAV}}[n] \hspace*{-0.5mm} + \hspace*{-0.5mm} P_{\mathrm{static}}\Bigg]\Delta_{\mathrm{T}} \hspace*{-0.5mm} \le \hspace*{-0.5mm}  q[n], \forall n, \notag \\[-2mm]
%%%%
&&\hspace*{-7mm}   \mbox{C2: } q[n+1] \hspace*{-0.5mm} \le \hspace*{-0.5mm} q[n] \hspace*{-0.5mm} + \hspace*{-0.5mm} \underline{P}^{\mathrm{solar}}\big(z[n]\big)\Delta_{\mathrm{T}} \hspace*{-0.5mm} - \hspace*{-0.5mm} \Bigg[ \overset{N_{\mathrm{F}}}{\underset{i = 1}{\sum}} \overset{K}{\underset{k = 1}{\sum}} \frac{1}{\varepsilon} \tilde{p}_k^i[n] \hspace*{-0.5mm}  + \hspace*{-0.5mm} P_{\mathrm{UAV}}[n] + P_{\mathrm{static}}\Bigg]\Delta_{\mathrm{T}},   \notag \\[-3mm]
%%%%%%%%%%%%%%%%%%%%%%%%%%%%%%%
&&\hspace*{-7mm}  \mbox{C5: } \tilde{p}_k^i[n] \hspace*{-0.5mm}\ge \hspace*{-0.5mm} 0, \forall i,k,n,
  \hspace*{28mm} \mbox{C11: } \overset{N_{\mathrm{F}}}{\underset{i = 1}{\sum}} \overset{K}{\underset{k = 1}{\sum}}  \tilde{p}_k^i[n] \hspace*{-0.5mm} \le \hspace*{-0.5mm} P_{\mathrm{max}}, \forall n, \notag \\[-5mm]
%%%%%%
&&\hspace*{-7mm}   \mbox{C14: } \overset{N_{\mathrm{F}}}{\underset{i = 1}{\sum}} \tilde{R}_k^i[n](\tilde{\mathbf{p}},\mathbf{r}) \hspace*{-0.5mm} \ge \hspace*{-0.5mm} R^{\mathrm{req}}_k, \forall k,n,
\hspace*{7mm}  \mbox{C15: } \norm{\mathbf{r}[n]\hspace*{-0.5mm}-\hspace*{-0.5mm}\mathbf{r}_k}^2 \le \theta_k[n], \forall k,n,   \notag
\end{eqnarray}
where $\theta_k[n]$ is an auxiliary variable and $\bm{\theta}\in\mathbb{R}^{K \times1}$ is the collection of all $\theta_{k}[n]$. For simplicity, the bandwidth $N_{\mathrm{F}}\mathcal{B}$ is omitted from the objective function.
We note that constraint C12 and binary selection constraint C13 have been absorbed into the objective function via the penalty term $\xi \sum_{j\neq k}^{K} H_k^i[n]\tilde{p}_j^i[n]$.
The problem formulations in \eqref{equiv-prob} and \eqref{prob} are equivalent when in \eqref{equiv-prob} on each subcarrier at most one of the powers $\tilde{p}_{k}^i[n]$ is non-zero in each time slot.
Now, we introduce the following theorem which confirms the equivalence of \eqref{equiv-prob} and \eqref{prob}.

\begin{Thm} \label{Thm-eqiv-prob}
For a sufficiently large $\xi \gg 1$,  the optimal subcarrier assignment strategy for maximizing the system sum throughput in \eqref{equiv-prob} assigns each subcarrier exclusively to at most one user in each time slot and no subcarrier is shared by multiple users. Hence, \eqref{prob} and \eqref{equiv-prob} are equivalent.
\end{Thm}
\emph{\quad Proof: } Please refer to the Appendix. \hfill\qed

Next, we note that constraints C1 and C2 in \eqref{equiv-prob} are non-convex and non-monotonic functions. To facilitate the use of monotonic optimization, we rewrite C1 and C2 in the following equivalent form:\vspace*{-3mm}
\begin{eqnarray}
&&\hspace*{-7mm}   \mbox{C1a: } \overset{N_{\mathrm{F}}}{\underset{i = 1}{\sum}} \overset{K}{\underset{k = 1}{\sum}} \frac{1}{\varepsilon}  \tilde{p}_k^i[n] \hspace*{-0.5mm}  + \hspace*{-0.5mm} P_{\mathrm{UAV}}[n] + P_{\mathrm{static}} \hspace*{-0.5mm} \le \hspace*{-0.5mm}  t[n],  \\
%%%%
&&\hspace*{-7mm}   \mbox{C1b: } t[n]\Delta_{\mathrm{T}}\hspace*{-0.5mm} \le \hspace*{-0.5mm}  q[n],  \\
%%%%
&&\hspace*{-7mm}   \mbox{C2: } q[n+1] - q[n] + t[n]\Delta_{\mathrm{T}} -C_2\Delta_{\mathrm{T}} \le \frac{C_1\Delta_{\mathrm{T}}}{1+e^{-k_{c}(z[n]-\alpha)}}, \label{Battery_storg_cons}
\end{eqnarray}
where $t[n]$ is an auxiliary optimization variable. Note that constraint C1a is non-convex due to the term $P_{\mathrm{UAV}}[n]$. To tackle this problem, we introduce the following equivalent transformation of constraint C1a:\vspace*{-2mm}
\begin{eqnarray}
&&\hspace*{-7mm} \overline{\mbox{C1}}\mbox{a: } \overset{N_{\mathrm{F}}}{\underset{i = 1}{\sum}} \overset{K}{\underset{k = 1}{\sum}} \frac{1}{\varepsilon} \tilde{p}_k^i[n] \hspace*{-0.5mm}  + \hspace*{-0.5mm}  \varrho_1 \mu[n] + W v_{z}[n] +  \varrho_2 \big(\overline{v}[n]\big)^{3} + P_{\mathrm{static}} \hspace*{-0.5mm} \le \hspace*{-0.5mm} t[n] ,  \\[-2mm]
% &&\hspace*{-7mm} \mbox{C1b: } t[n] \le q[n],  \\
&&\hspace*{-7mm} \mbox{C16: }  \mu[n] \ge \frac{1}{\sqrt{\norm{(v_x[n], v_y[n])}^2 \hspace*{-0.5mm} + \hspace*{-0.5mm} \sqrt{\norm{(v_x[n], v_y[n])}^4 \hspace*{-0.5mm} + \hspace*{-0.5mm} 4 V_{\mathrm{h}}^4}}},  \\
&&\hspace*{-7mm} \mbox{C17: }  \norm{(v_x[n], v_y[n])} \le \overline{v}[n],
\end{eqnarray}
where $\mu[n]$ and $\overline{v}[n]$ are auxiliary optimization variables. We note that $\overline{\mbox{C1}}\mbox{a}$ and C17 are convex constraints and C16 is monotonically increasing with $\mu[n]$. In addition, constraint C2 in \eqref{Battery_storg_cons} is non-convex due to the logistic function term $\frac{C_1\Delta_{\mathrm{T}}}{1+e^{-k_{c}(z[n]-\alpha)}}$. To overcome this difficulty, we take the logarithm of both sides of $\mbox{C2}$ which leads to constraint $\overline{\mbox{C2}}$:\vspace*{-2mm}
\begin{eqnarray} \label{diff_log_cons}
&&\hspace*{-7mm}   \overline{\mbox{C2}}\mbox{: } \ln\big(q[n+1] - q[n] + t[n]\Delta_{\mathrm{T}} -C_2\Delta_{\mathrm{T}}\big) - \ln\Big(\frac{C_1\Delta_{\mathrm{T}}}{1+e^{-k_{c}(z[n]-\alpha)}}\Big) \le 0.
\end{eqnarray}
In \eqref{diff_log_cons}, the term $\ln\Big(\frac{C_1\Delta_{\mathrm{T}}}{1+e^{-k_{c}(z[n]-\alpha)}}\Big)$ is a concave function, thereby constraint $\overline{\mbox{C2}}$ is the difference of two concave logarithmic functions which is still non-convex. To circumvent this issue, we replace $\overline{\mbox{C2}}$ with the following equivalent constraints:\vspace*{-2mm}
\begin{eqnarray}
&&\hspace*{-7mm}   \overline{\mbox{C2}}\mbox{a: } \ln(\varpi[n]) + \tau[n] \le E,  \\[-2mm]
&&\hspace*{-7mm}   \overline{\mbox{C2}}\mbox{b: } \ln\Big(\frac{C_1\Delta_{\mathrm{T}}}{1+e^{-k_{c}(z[n]-\alpha)}}\Big) + \tau[n] \ge E,  \\[-2mm]
&&\hspace*{-7mm}   \mbox{C18: } 0 \le q[n+1] - q[n] + t[n]\Delta_{\mathrm{T}} -C_2\Delta_{\mathrm{T}} \le \varpi[n],
\end{eqnarray}
where $\varpi[n]$ and $\tau[n]$ are auxiliary optimization variables and $E = \ln\Big(\frac{C_1\Delta_{\mathrm{T}}}{1+e^{-k_{c}(z_{\mathrm{max}}-\alpha)}}\Big)$ is a constant. We note that constraint $\overline{\mbox{C2}}\mbox{a}$ is a monotonically increasing function in $\varpi[n]$ and $\tau[n]$ and $\overline{\mbox{C2}}\mbox{b}$ and C18 are convex constraints.

Then, to facilitate the application of monotonic optimization theory, we define auxiliary variable $\chi_{k}^{i}[n]$ which satisfies the following constraint: \vspace*{-1mm}
\begin{eqnarray}
&&\hspace*{-7mm}   \mbox{C19: } 1 \le \chi_{k}^{i}[n] \le \frac{f_{k}^{i}[n](\tilde{\mathbf{p}},\bm{\theta})}{g_{k}^{i}[n](\tilde{\mathbf{p}},\bm{\theta})},
\end{eqnarray}
where  \vspace*{-3mm}
%%%%%%%%%%%%%%%%%%%%%%%%%%%%%%%%%%%%%%%%%%%%%%%%
\begin{eqnarray}
\hspace*{-5mm}f_{k}^{i}[n](\tilde{\mathbf{p}},\bm{\theta}) \hspace*{-2mm}&=&\hspace*{-2mm} H_k^i[n] \tilde{p}_k^i[n] \hspace*{-1mm} + \hspace*{-1mm} \xi \sum_{j \neq k}^{K} H_k^i[n]\tilde{p}_j^i[n] \hspace*{-1mm} + \hspace*{-1mm} \theta_k[n] \,\,\, \text{and} \,\,\,
%%%%%%%
g_{k}^{i}[n](\tilde{\mathbf{p}},\bm{\theta})\hspace*{-1mm}=
\hspace*{-1mm} \xi \sum_{j \neq k}^{K} H_k^i[n]\tilde{p}_j^i[n] \hspace*{-1mm} + \hspace*{-1mm} \theta_k[n],
\end{eqnarray}
%%%%%%%%%%%%%%%%%%%%%%%%%%%%%%%%%%%%%%%%%%%%%%%%%
capture the numerator and the denominator inside the logarithmic function of the objective function in \eqref{equiv-prob}, respectively.
Therefore, the original problem in \eqref{equiv-prob} can be equivalently rewritten as:\vspace*{-3mm}
\begin{eqnarray}\label{MO-pro}
\hspace*{-1mm} \underset{\bm{\chi},\bm{\mu},\bm{\varpi},\bm{\tau}}{\maxo}\hspace*{-1mm} && \hspace*{-1mm} \overset{N_{\mathrm{T}}}{\underset{n = 1}{\sum}} \overset{N_{\mathrm{F}}}{\underset{i = 1}{\sum}} \overset{K}{\underset{k = 1}{\sum}} \log_2(\chi_{k}^{i}[n])  \quad \quad \mbox{s.t.}\hspace*{1mm}
 (\bm{\chi},\bm{\mu},\bm{\varpi},\bm{\tau})\in\mathcal{V},
\end{eqnarray}
where  $\bm{\chi}\in\mathbb{R}^{N_{\mathrm{T}}N_{\mathrm{F}}K\times 1}$, $\bm{\mu}\in\mathbb{R}^{N_{\mathrm{T}}\times 1}$, $\bm{\varpi}\in\mathbb{R}^{N_{\mathrm{T}}\times 1}$, and $\bm{\tau}\in\mathbb{R}^{N_{\mathrm{T}}\times 1}$ are the collections of all $\chi_{k}^{i}[n]$, $\mu[n]$, $\varpi[n]$, and $\tau[n]$, respectively, and $\mathcal{V}=\mathcal{G}\cap\mathcal{H}$ is the feasible set. In particular, $\mathcal{G}$ is a normal set and $\mathcal{H}$ is a conormal set \cite{tuy2000monotonic,zhang2013monotonic}, and they are given by \vspace*{-2mm}
\begin{eqnarray}
\hspace*{-3mm}\mathcal{G}\hspace*{-3mm}&=&\hspace*{-3mm}\Big\{ (\bm{\chi},\bm{\mu},\bm{\varpi},\bm{\tau}) \mid
(\bm{\chi},\bm{\mu},\bm{\varpi},\bm{\tau}) \in \mathcal{P}\Big\} \quad \text{and} \quad
\mathcal{H}\hspace*{-1mm}=\hspace*{-1mm}\Big\{ (\bm{\chi},\bm{\mu}) \mid (\bm{\chi},\bm{\mu})\in\mathcal{Q}\Big\},
\end{eqnarray}
where feasible set $\mathcal{P}$ is spanned by constraints $\overline{\mbox{C1}}\mbox{a}$, $\mbox{C1b}$, $\overline{\mbox{C2}}\mbox{a}$, $\overline{\mbox{C2}}\mbox{b}$, $\mbox{C3--C11}$, $\mbox{C15}$, and $\mbox{C17--C19}$ and feasible set $\mathcal{Q}$ is spanned by constraints $\mbox{C14}$ and C16.

\begin{table}\vspace*{-8mm}
%\caption{Outer Polyblock Approximation Algorithm.}\label{table:algorithm}\vspace*{-0cm}
%\small
\begin{algorithm} [H]                    % enter the algorithm environment
\caption{Sequential Polyblock Approximation Algorithm}          % give the algorithm a caption
\label{alg1}                           % and a label for \ref{} commands later in the document
\begin{algorithmic} [1]
\small          % enter the algorithmic environment
%\STATE Initialize polyblock $\mathcal{D}^{(1)}$ with vertex set $\bm{\Upsilon}^{(1)}\hspace*{-0.5mm} =\hspace*{-0.5mm}\bm{\upsilon}^{(1)} $ where the elements of $\bm{\upsilon}^{(1)}$ are set as:
\STATE Initialize polyblock $\mathcal{D}^{(1)}$. The vertex $\bm{\upsilon}^{(1)}\hspace*{-0.5mm}=\hspace*{-0.5mm} \big(\bm{\chi}^{(1)}\hspace*{-0.5mm},\hspace*{-0.5mm}\bm{\mu}^{(1)} \hspace*{-0.5mm},\hspace*{-0.5mm}\bm{\varpi}^{(1)}\hspace*{-0.5mm},\hspace*{-0.5mm}\bm{\tau}^{(1)}\big)$ is initialized by setting its elements as follows:
$\chi_{k}^{i}[n]=1 + H_{k}^{i}[n] P_{\mathrm{max}}$, $\mu[n] = 1/(\sqrt{2}V_{\mathrm{h}})$, $\varpi[n] = e^{E}$, and $\tau[n]=E$, $\forall k,i,n.$

\STATE Set error tolerance $\epsilon_1 \ll 1$ and iteration index $m=1$

\REPEAT [Main Loop]
%%%%%%%%%%%%%%%%%%%%
\STATE Calculate the projection of vertex $\bm{\upsilon}^{(m)}$ onto set $\mathcal{G}$ , i.e., $\bm{\Phi}(\bm{\upsilon}^{(m)})$, via $\textbf{Algorithm 2}$

\STATE Generate $D$ new vertices $\tilde{\bm{\Upsilon}}^{(m)} \hspace*{-0.5mm}=\hspace*{-0.5mm} \big\{\hspace*{-0.5mm}\tilde{\bm{\upsilon}}^{(m)}_{1}\hspace*{-0.8mm},\ldots,\hspace*{-0.5mm}\tilde{\bm{\upsilon}}^{(m)}_{D}\hspace*{-0.5mm}\big\}$, where $\tilde{\bm{\upsilon}}^{(m)}_{j} \hspace*{-0.5mm} = \hspace*{-0.5mm} \bm{\upsilon}^{(m)} \hspace*{-0.5mm} - \hspace*{-0.5mm} \big(\upsilon^{(m)}_j \hspace*{-0.5mm} - \hspace*{-0.3mm} \phi_j(\bm{\upsilon}^{(m)})\big)\mathbf{u}_j$, $j \hspace*{-0.5mm} \in \hspace*{-0.5mm}\{1\hspace*{-0.2mm},\ldots,\hspace*{-0.5mm}D\}$

\STATE Construct a smaller polyblock $\mathcal{D}^{(m+1)}$ with vertex set $\bm{\Upsilon}^{{(m+1)}} = \big(\bm{\Upsilon}^{(m)} -\bm{\upsilon}^{(m)}\big) \cup \tilde{\bm{\Upsilon}}^{(m)}$

\STATE  Find $\bm{\upsilon}^{(m+1)}$ as that vertex of $\bm{\Upsilon}^{{(m+1)}} \cap \mathcal{H}$ whose projection maximizes the objective function of the problem, i.e., $\bm{\upsilon}^{(m+1)}=\underset{\bm{\upsilon} \in\bm{\Upsilon}^{(m+1)}\cap \mathcal{H}}{\arg \max} \Big\{ \overset{N_{\mathrm{T}}}{\underset{n = 1}{\sum}} \overset{N_{\mathrm{F}}}{\underset{i = 1}{\sum}} \overset{K}{\underset{k = 1}{\sum}} \log_2(\chi_{k}^{i}[n]) \Big\}$, and set $m = m+1$

\UNTIL $\frac{\norm{\bm{\upsilon}^{(m)} -\mathbf{\Phi}(\bm{\upsilon}^{(m)})}} {\norm{\bm{\upsilon}^{(m)}}} \le \epsilon_1$

\STATE $\bm{\upsilon}^{*} =\mathbf{\Phi}(\bm{\upsilon}^{(m)})$ and $(\tilde{\mathbf{p}}^*,\mathbf{r}^*,\mathbf{v}^*,\mathbf{q}^*,\bm{\theta}^*)$ are obtained when calculating $\mathbf{\Phi}(\bm{\upsilon}^{(m)})$
\end{algorithmic}
\end{algorithm}\vspace*{-8mm}
\end{table}

\begin{figure}
\centering \vspace*{-4mm}
  \subfigure[]{
    \label{fig:polyblock1} %% label for first subfigure
    \begin{minipage}[b]{0.22\textwidth}
      \centering \vspace*{-3mm}
      \includegraphics[width=1.5in]{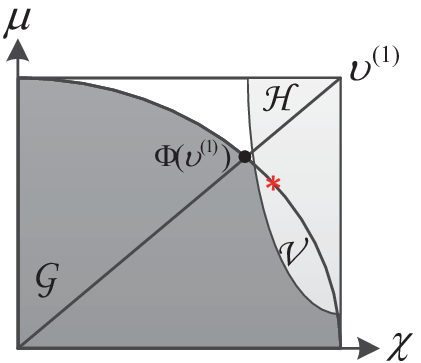}\vspace*{-3mm}
    \end{minipage}}%
  \subfigure[]{
    \label{fig:polyblock2}
    \begin{minipage}[b]{0.22\textwidth}
    %%%%%%%%%%%%%
      \centering \vspace*{-3mm}
      \includegraphics[width=1.5in]{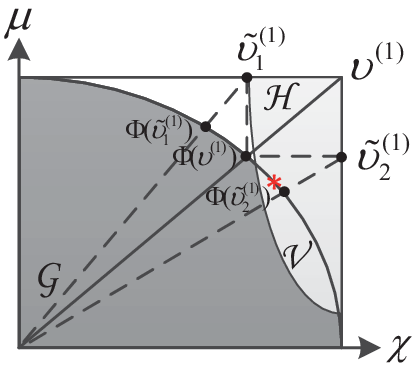}\vspace*{-3mm}
    \end{minipage}}
    \subfigure[]{
    \label{fig:polyblock3}
    \begin{minipage}[b]{0.22\textwidth}
    %%%%%%%%%%%%%%
      \centering \vspace*{-3mm}
      \includegraphics[width=1.5in]{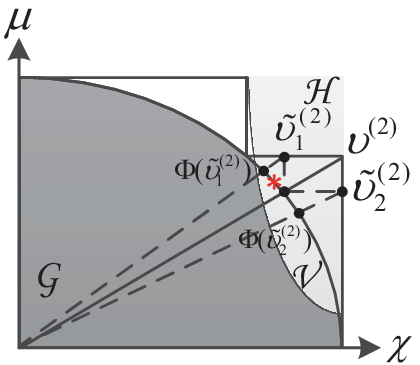}\vspace*{-3mm}
    \end{minipage}}%
    \subfigure[]{
    \label{fig:polyblock4}
    \begin{minipage}[b]{0.22\textwidth}
    %%%%%%%%%%%%%%
      \centering \vspace*{-3mm}
      \includegraphics[width=1.5in]{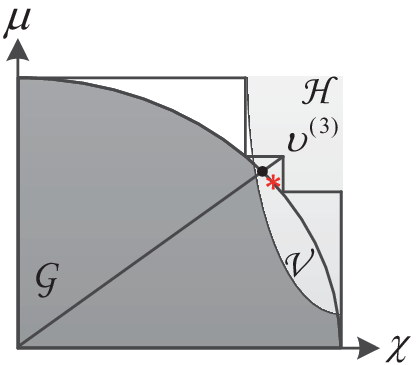}\vspace*{-3mm}
    \end{minipage}} \vspace*{-3mm}
  \caption{Illustration of the sequential polyblock approximation algorithm. The red star is the optimal point on the upper boundary of the feasible set $\mathcal{V}$.}
  \label{fig:polyblock}  \vspace*{-7mm}
\end{figure}

In \eqref{MO-pro}, the objective function is a monotonically increasing function. The constraint functions are monotonically increasing functions or convex functions and determine the feasible set $\mathcal{V}$, which is the intersection of the normal set $\mathcal{G}$ and the conormal set $\mathcal{H}$. Hence, problem \eqref{MO-pro} is in the canonical form of a monotonic optimization problem \cite{tuy2000monotonic,zhang2013monotonic}. According to monotonic optimization theory \cite{tuy2000monotonic,zhang2013monotonic}, the optimal solution of \eqref{MO-pro} lies on the upper boundary of the feasible set $\mathcal{V}$ and can be approached via the sequential polyblock approximation \cite{sun2016optimalJournal}.
%%%%%%%%%%%%%%%%%%%%%%%%%%%%
In particular, first, we initialize a polyblock $\mathcal{D}^{(1)}$ that encloses the feasible set $\mathcal{V}=\mathcal{G} \cap  \mathcal{H}$.
The vertex set of $\mathcal{D}^{(1)}$ is denoted as $\bm{\Upsilon}^{(1)}$ and contains one vertex $\bm{\upsilon}^{(1)}$. Here, vertex $\bm{\upsilon}^{(1)}$ is defined as $\bm{\upsilon}^{(1)}\triangleq (\bm{\chi}^{(1)},\bm{\mu}^{(1)},\bm{\varpi}^{(1)},\bm{\tau}^{(1)})$ and represents the optimization variables in \eqref{MO-pro}.
%%%%
Based on vertex $\bm{\upsilon}^{(1)}$, we generate $D = N_{\mathrm{T}}(N_{\mathrm{F}}K + 3)$ new vertices $\tilde{\bm{\Upsilon}}^{(1)} =\big\{\tilde{\bm{\upsilon}}^{(1)}_{1},\ldots,\tilde{\bm{\upsilon}}^{(1)}_{D}\big\}$. Specifically,
$\tilde{\bm{\upsilon}}^{(1)}_{j}\hspace*{-0.6mm} =\hspace*{-0.6mm}\bm{\upsilon}^{(1)} \hspace*{-0.6mm}-\hspace*{-0.6mm}\big(\upsilon^{(1)}_j \hspace*{-0.6mm} - \hspace*{-0.6mm} \phi_j(\bm{\upsilon}^{(1)})\big)\mathbf{u}_j$, $j\in\{1,\hspace*{-0.5mm}\ldots, \hspace*{-0.5mm}D\}$, where $\upsilon^{(1)}_j$ and $\phi_j(\bm{\upsilon}^{(1)})$ are the $j$-th elements of $\bm{\upsilon}^{(1)}$ and $\mathbf{\Phi}(\bm{\upsilon}^{(1)})$, respectively. Here, $\mathbf{\Phi}(\bm{\upsilon}^{(1)})\in \mathbb{C}^{D \times 1}$ is the projection of $\bm{\upsilon}^{(1)}$ onto set $\mathcal{G}$,  and $\mathbf{u}_j$ is a unit vector containing only one non-zero element at position $j$.
%%%%%%
Then, we shrink $\mathcal{D}^{(1)}$ by replacing $\bm{\upsilon}^{(1)}$ with the $D$ vertices $\tilde{\bm{\Upsilon}}^{(1)}$, leading to a new polyblock $\mathcal{D}^{(2)}$ with vertex set $\bm{\Upsilon}^{(2)}\hspace*{-0.5mm} =\hspace*{-0.5mm}\big(\bm{\Upsilon}^{(1)} \hspace*{-0.5mm}-\hspace*{-0.5mm}\bm{\upsilon}^{(1)}\big) \hspace*{-0.5mm} \cup \hspace*{-0.5mm} \tilde{\bm{\Upsilon}}^{(1)}$.
The new polyblock $\mathcal{D}^{(2)}$ is smaller than $\mathcal{D}^{(1)}$, but still contains the feasible set $\mathcal{V}$.
%%%%%%%
Then, we choose $\bm{\upsilon}^{(2)}$ as the optimal vertex of $\bm{\Upsilon}^{(2)} \hspace*{-0.5mm} \cap \hspace*{-0.5mm} \mathcal{H}$ whose projection maximizes the objective function of the problem in \eqref{MO-pro}, i.e., $\bm{\upsilon}^{(2)}\hspace*{-0.5mm} =\hspace*{-0.5mm}\underset{\bm{\upsilon} \in\bm{\Upsilon}^{(2)}\cap \mathcal{H}}{\arg \max} \Big\{ \hspace*{-0.5mm} \overset{N_{\mathrm{T}}}{\underset{n = 1}{\sum}} \overset{N_{\mathrm{F}}}{\underset{i = 1}{\sum}} \overset{K}{\underset{k = 1}{\sum}} \log_2(\chi_{k}^{i}[n])\hspace*{-0.5mm}  \Big\}$. Similarly, we repeat the above procedure to shrink $\mathcal{D}^{(2)}$ based on $\bm{\upsilon}^{(2)}$, constructing a smaller polyblock and so on, i.e., $\mathcal{D}^{(1)} \hspace*{-1mm} \supset \hspace*{-1mm} \mathcal{D}^{{(2)}} \hspace*{-1mm} \supset \hspace*{-1mm} \dots \hspace*{-1mm} \supset \hspace*{-1mm}  \mathcal{V}$. The algorithm terminates if  $\frac{\norm{\bm{\upsilon}^{(m)} -\mathbf{\Phi}(\bm{\upsilon}^{(m)})}} {\norm{\bm{\upsilon}^{(m)}}} \hspace*{-0.6mm} \le \hspace*{-0.6mm} \epsilon_1$, where the error tolerance constant $\epsilon_1 \hspace*{-0.6mm} > \hspace*{-0.6mm} 0$ specifies the accuracy of the approximation.
Figure \ref{fig:polyblock} illustrates the algorithm for a simple case, which, for simplicity of presentation, includes only two optimization variables, i.e., $\chi$ and $\mu$.
We summarize the proposed sequential polyblock approximation algorithm in \textbf{Algorithm 1}.

The projection of vertex $\bm{\upsilon}^{(m)}$, i.e., $\bm{\Phi}(\bm{\upsilon}^{(m)})$, is required in each iteration of \textbf{Algorithm 1}.
Specifically, $\bm{\Phi}(\bm{\upsilon}^{(m)})$ can be represented as $\bm{\Phi}(\bm{\upsilon}^{(m)})=\lambda \bm{\upsilon}^{(m)} = \lambda \big(\bm{\chi}^{(m)},\bm{\mu}^{(m)} ,\bm{\varpi}^{(m)},\bm{\tau}^{(m)}\big)$, where $\lambda=\max\{ \beta  \mid \beta \bm{\upsilon}^{(k)}  \in \mathcal{G}\}$ is the projection parameter and the value of $\lambda$ is between $0$ and $1$, i.e., $\lambda \in [0,1]$ \cite{tuy2000monotonic,zhang2013monotonic}. Thus, we can obtain $\lambda$ via the bisection search method \cite{book:convex}.
In particular, for a given projection parameter $\overline{\lambda}$ and vertex $\bm{\upsilon}^{(m)}$, we can check the feasibility of $\overline{\lambda}\bm{\upsilon}^{(m)} \in  \mathcal{G}$ by checking the feasibility of the following convex problem:\vspace*{-2mm}
\begin{eqnarray} \label{lambda_feasib}
&& \hspace*{-1mm} \underset{\tilde{\mathbf{p}},\mathbf{r},\mathbf{v},\overline{\mathbf{v}},\mathbf{q},\mathbf{t},\bm{\theta}}{\maxo} \quad 1   \notag \\
\hspace*{-1mm}\mbox{s.t.}\hspace*{1mm}
&&\hspace*{-7mm}  \overline{\mbox{C1}}\mbox{a: } \overset{N_{\mathrm{F}}}{\underset{i = 1}{\sum}} \overset{K}{\underset{k = 1}{\sum}} \frac{1}{\varepsilon} \tilde{p}_k^i[n] \hspace*{-0.5mm}  + \hspace*{-0.5mm}  \varrho_1 \overline{\lambda} \big(\mu[n]\big)^{(m)} + W v_{z}[n] +  \varrho_2 \big(\overline{v}[n]\big)^{3} + P_{\mathrm{static}} \hspace*{-0.5mm} \le \hspace*{-0.5mm} t[n], \forall n, \notag \\
&&\hspace*{-7mm}   \overline{\mbox{C2}}\mbox{a: } \ln\Big(\overline{\lambda}\big(\varpi[n]\big)^{(m)}\Big) + \overline{\lambda}\big(\tau[n]\big)^{(m)} \le E, \notag \\
&&\hspace*{-7mm}   \overline{\mbox{C2}}\mbox{b: } \ln\Big(\frac{C_1\Delta_{\mathrm{T}}}{1+e^{-k_{c}(z[n]-\alpha)}}\Big) + \overline{\lambda}\big(\tau[n]\big)^{(m)} \ge E, \forall n, \notag \\
&&\hspace*{-7mm}   \mbox{C18: } 0 \le q[n+1] - q[n] + t[n]\Delta_{\mathrm{T}} -C_2\Delta_{\mathrm{T}} \le \overline{\lambda} \big(\varpi[n]\big)^{(m)}, \forall n, \notag \\
&&\hspace*{-7mm}   \mbox{C19: } \overline{\lambda} \chi_d^{(m)} g_d(\tilde{\mathbf{p}},\mathbf{r}) - f_d(\tilde{\mathbf{p}},\mathbf{r}) \le 0, \forall d, \hspace*{12mm}  \mbox{C1b}, \mbox{C3--C11, C15, C17},
\end{eqnarray}
where $\overline{\mathbf{v}} \in\mathbb{R}^{N_{\mathrm{T}}\times 1}$ and $\mathbf{t} \in\mathbb{R}^{N_{\mathrm{T}}\times 1}$ are the collections of all $\overline{v}[n]$ and $t[n]$, respectively. The constraints in \eqref{lambda_feasib} span the feasible set $\mathcal{G}$. We summarize the proposed projection calculation algorithm in \textbf{Algorithm 2}. We note that problem \eqref{lambda_feasib} can be solved efficiently by standard convex optimization solvers such as CVX \cite{website:CVX}. Besides, the optimal resource allocation policy $\{\tilde{\mathbf{p}},\mathbf{r},\mathbf{v},\mathbf{q}\}$ is obtained from \textbf{Algorithm 2}. In addition, we can recover the optimal subcarrier allocation policy $\mathbf{s}$ from the obtained $\tilde{\mathbf{p}}$ by allocating $s_{k}^i[n] = 1$ if $\tilde{p}_k^i[n] > 0$ and $s_{k}^i[n] = 0$ if $\tilde{p}_k^i[n] = 0$.

\begin{table}[t]\vspace*{-5mm}
\begin{algorithm} [H]                    % enter the algorithm environment
\caption{Optimal Projection via Bisection Search}          % give the algorithm a caption
\label{alg2}                           % and a label for \ref{} commands later in the document
\begin{algorithmic} [1]
\small          % enter the algorithmic environment
\STATE Initialize $\lambda_{\mathrm{min}}=0$ and $\lambda_{\mathrm{max}}=1$ and set the error tolerance $\epsilon_2 \ll 1$. \vspace*{-2mm}

\REPEAT\vspace*{-0mm}
%%%%%%%%%%%%%%%%%%%%
\STATE Set $\overline{\lambda} = (\lambda_{\mathrm{min}}+ \lambda_{\mathrm{max}})/2$.

\STATE Check the feasibility of $\overline{\lambda}$, i.e., whether $\overline{\lambda}\bm{\upsilon}^{(m)} \hspace*{-0.8mm} \in \hspace*{-0.8mm} \mathcal{G}$, by solving \eqref{lambda_feasib}. If \eqref{lambda_feasib} is feasible, $\lambda_{\mathrm{min}} \hspace*{-0.6mm} = \hspace*{-0.6mm} \overline{\lambda}$; else $\lambda_{\mathrm{max}} \hspace*{-0.6mm} = \hspace*{-0.6mm} \overline{\lambda}$.

\UNTIL $\lambda_{\mathrm{max}}-\lambda_{\mathrm{min}} \le \epsilon_2$.

\STATE $\lambda=\lambda_{\mathrm{min}}$ and the projection is $\mathbf{\Phi}(\bm{\upsilon}^{(m)}) =\lambda \bm{\upsilon}^{(m)}$.
The corresponding resource allocation policy $\{\tilde{\mathbf{p}},\mathbf{r},\mathbf{v},\mathbf{q}\}$ is obtained by solving \eqref{lambda_feasib} for $\overline{\lambda}=\lambda_{\mathrm{min}}$.

\end{algorithmic}
\end{algorithm}\vspace*{-14mm}
\end{table}

We note that although the proposed optimal offline resource allocation algorithm requires non-causal knowledge of the channel gains, the obtained optimal performance can serve as a benchmark for any optimal and suboptimal offline and online resource allocation scheme.
Besides, the proposed monotonic optimization based optimal trajectory and resource allocation algorithm serves as a building block
for designing the optimal online resource allocation policy.
In the next section, we will study online trajectory design and resource allocation schemes which require only causal knowledge of the channel states.

\vspace*{-3mm}
\section{Online Trajectory and Resource Allocation Design}
In this section, we study online trajectory and resource allocation designs which require only causal knowledge of the channel states.  The considered online resource allocation is formulated as a non-convex optimization problem.
Inspired by the derived offline solution, we develop an optimal online resource allocation algorithm.
Besides, a low-complexity suboptimal scheme is also proposed.

\vspace*{-3mm}
\subsection{Achievable and Expected Data Rate}
The online trajectory and resource allocation is performed in each time slot. Assume that the index of the current time slot is $n_0$, where $1 \le n_0 \le N_{\mathrm{T}}$, and subcarrier $i$ is allocated to user $k$ in time slot $n_0$. The UAV can acquire near-perfect channel state information (CSI) of user $k$ via handshaking between the UAV and user $k$ at the beginning of time slot $n_0$. Then, with the variables defined in the previous sections, the achievable data rate (bits/s) on subcarrier $i$ in current time slot $n_0$ is given by: \vspace*{-2mm}
\begin{equation}\label{current-rate_k}
{R}_k^i[n_0](\mathbf{p},\mathbf{s},\mathbf{r})=s_{k}^i[n_0] \mathcal{B} \log_2 \Big( 1 + \frac{H_k^i[n_0] p_k^i[n_0]}{\norm{\mathbf{r}[n_0]-\mathbf{r}_k}^2} \Big).
\end{equation}
%%%%%%%%%%%%%%%%%%%%
Besides, for future time slot $n$, i.e., $n_0 < n \le N_{\mathrm{T}}$, the CSI of the users is not available at the UAV, yet. Hence, we employ the expected data rate for online trajectory and resource allocation design.
In particular, assuming that subcarrier $i$ is allocated to user $k$ in time slot $n$, the expected data rate on subcarrier $i$ is given by: \vspace*{-3mm}
\begin{eqnarray}\label{expected-rate_k}
{E\hspace*{-0.5mm}R}_k^i[n](\mathbf{p},\mathbf{s},\mathbf{r}) \hspace*{-2mm} &=& \hspace*{-2mm} \mathcal{E} \Big\{s_{k}^i[n] \mathcal{B}\log_2 \Big( 1 + \frac{H_k^i[n] p_k^i[n]}{\norm{\mathbf{r}[n]-\mathbf{r}_k}^2} \Big) \Big\},
\end{eqnarray}
for  $n_0 < n \le N_{\mathrm{T}}$.
For facilitating a tractable resource allocation algorithm design, we rewrite the expected data rate as \cite{teh2007collapsed}: \vspace*{-3mm}
\begin{eqnarray}\label{upperbound-rate_k}
&&\hspace*{-6mm}{E\hspace*{-0.5mm}R}_k^i[n](\mathbf{p},\mathbf{s},\mathbf{r}) \notag \\
\hspace*{-2mm} &\overset{(a)}{=}& \hspace*{-2mm}  s_{k}^i[n] \mathcal{B} \,\, \mathcal{E}\Bigg\{ \log_2 (1 + \mathcal{E} \{H_k^i[n]\} \delta ) + \frac{ H_k^i[n] \delta- \mathcal{E} \{H_k^i[n]\} \delta}{\big(1+\mathcal{E} \{H_k^i[n]\} \delta\big) \ln 2} - \frac{\big(H_k^i[n] \delta- \mathcal{E} \{H_k^i[n]\} \delta\big)^2}{2\ln 2\big(1+\mathcal{E} \{H_k^i[n]\} \delta \big)^2} + o\Bigg\} \notag \\
%%%%%%
\hspace*{-2mm} &=& \hspace*{-2mm} s_{k}^i[n] \mathcal{B}\Bigg[ \log_2 \Big( 1+\mathcal{E} \{H_k^i[n]\} \delta \Big) \hspace*{-1mm}- \hspace*{-1mm} \frac{\frac{\zeta^2 {\delta}^2}{{\mathcal{N}_0^2 \mathcal{B}^2}}\mathrm{Var}\{\abs{h_k^i[n]}^2\}}{2\ln 2\big(1\hspace*{-1mm}+\hspace*{-1mm}\frac{\zeta \delta}{{\mathcal{N}_0 \mathcal{B}}}\mathcal{E}\{\abs{h_k^i[n]}^2\}\big)^2} \Bigg] + o \notag\\
\hspace*{-2mm}&=& \hspace*{-2mm} \overline{{E\hspace*{-0.5mm}R}}_k^i[n](\mathbf{p},\mathbf{s},\mathbf{r}) - s_{k}^i[n] \mathcal{B}\frac{\frac{\zeta^2 {\delta}^2}{{\mathcal{N}_0^2 \mathcal{B}^2}}\mathrm{Var}\{\abs{h_k^i[n]}^2\}}{2\ln 2\big(1\hspace*{-1mm}+\hspace*{-1mm}\frac{\zeta \delta}{{\mathcal{N}_0 \mathcal{B}}}\mathcal{E}\{\abs{h_k^i[n]}^2\}\big)^2} + o,
%%%%%%%%%%%%%%
%\hspace*{-2mm} &=& \hspace*{-2mm} s_{k}^i[n] \mathcal{B}\Bigg[ \log_2 \Big( 1+\mathcal{E} \{H_k^i[n]\} \delta \Big) \hspace*{-1mm}- \hspace*{-1mm} \frac{\frac{\zeta^2 {\delta}^2}{{\mathcal{N}_0^2 \mathcal{B}^2}}\mathrm{Var}\{\abs{h_k^i[n]}^2\}}{2(1\hspace*{-1mm}+\hspace*{-1mm}\frac{\zeta \delta}{{\mathcal{N}_0 \mathcal{B}}}\mathcal{E}\{\abs{h_k^i[n]}^2\})^2} + o\Big(\frac{\zeta^2 {\delta}^2}{{\mathcal{N}_0^2 \mathcal{B}^2}}\mathrm{Var}\{\abs{h_k^i[n]}^2\}\Big)\Bigg]\notag\\[-0mm]
%\hspace*{-2mm}&=& \hspace*{-2mm} \overline{{E\hspace*{-0.5mm}R}}_k^i[n](\mathbf{p},\mathbf{s},\mathbf{r}) - s_{k}^i[n] \mathcal{B}\Bigg[\frac{\frac{\zeta^2 {\delta}^2}{{\mathcal{N}_0^2 \mathcal{B}^2}}\mathrm{Var}\{\abs{h_k^i[n]}^2\}}{2(1\hspace*{-1mm}+\hspace*{-1mm}\frac{\zeta \delta}{{\mathcal{N}_0 \mathcal{B}}}\mathcal{E}\{\abs{h_k^i[n]}^2\})^2} - o\Big(\frac{\zeta^2 {\delta}^2}{{\mathcal{N}_0^2 \mathcal{B}^2}}\mathrm{Var}\{\abs{h_k^i[n]}^2\}\Big)\Bigg],
\end{eqnarray}
where ${\delta}=\frac{p_k^i[n]}{\norm{\mathbf{r}[n]-\mathbf{r}_k}^2}$, $(a)$ is due to the second-order Taylor series expansion of $\log_2(1+ H_k^i[n] \delta)$ around $\log_2(1+\mathcal{E} \{H_k^i[n]\} \delta)$, $o \ge 0$ denotes the higher-order infinitesimal terms, and we define
\begin{eqnarray}
\overline{{E\hspace*{-0.5mm}R}}_k^i[n](\mathbf{p},\mathbf{s},\mathbf{r}) = s_{k}^i[n] \mathcal{B}\log_2 \Big( 1 + \frac{\mathcal{E}\{H_k^i[n]\} p_k^i[n]}{\norm{\mathbf{r}[n]-\mathbf{r}_k}^2} \Big).
\end{eqnarray}
We note that $H_k^i[n]=\frac{\zeta}{{\mathcal{N}_0 \mathcal{B}}}\abs{h_k^i[n]}^2$ and $\mathcal{E}\{H_k^i[n]\} = \frac{\zeta}{{\mathcal{N}_0 \mathcal{B}}}\mathcal{E}\{\abs{h_k^i[n]}^2\}$ which is a constant and its value can be estimated by averaging historical observations $H_k^i[n]$. The difference between $\overline{{E\hspace*{-0.5mm}R}}_k^i[n](\mathbf{p},\mathbf{s},\mathbf{r})$ and ${E\hspace*{-0.5mm}R}_k^i[n](\mathbf{p},\mathbf{s},\mathbf{r})$ is bounded from above by
\begin{eqnarray}\label{bound_diff}
\overline{{E\hspace*{-0.5mm}R}}_k^i[n](\mathbf{p},\mathbf{s},\mathbf{r}) - {E\hspace*{-0.5mm}R}_k^i[n](\mathbf{p},\mathbf{s},\mathbf{r}) \hspace*{-2mm}&=&\hspace*{-2mm} s_{k}^i[n] \mathcal{B}\frac{\frac{\zeta^2 {\delta}^2}{{\mathcal{N}_0^2 \mathcal{B}^2}}\mathrm{Var}\{\abs{h_k^i[n]}^2\}}{2\ln 2\big(1\hspace*{-1mm}+\hspace*{-1mm}\frac{\zeta \delta}{{\mathcal{N}_0 \mathcal{B}}}\mathcal{E}\{\abs{h_k^i[n]}^2\}\big)^2} - o\notag \\[-0mm]
\hspace*{-2mm}&\le&\hspace*{-2mm} s_{k}^i[n] \mathcal{B}\frac{\mathrm{Var}\{\abs{h_k^i[n]}^2\}} {2\ln 2\big(\frac{{\mathcal{N}_0 \mathcal{B}} \norm{\mathbf{r}[n]-\mathbf{r}_k}^2}{\zeta p_k^i[n]} + \mathcal{E}\{\abs{h_k^i[n]}^2\}\big)^2} \notag \\[-0mm]
\hspace*{-2mm}&\le&\hspace*{-2mm} s_{k}^i[n] \mathcal{B}\frac{\mathrm{Var}\{\abs{h_k^i[n]}^2\}} {2 \ln 2 \big(\mathcal{E}\{\abs{h_k^i[n]}^2\}\big)^2}.
\end{eqnarray}
We note that the bound on the difference in \eqref{bound_diff} can be very small\footnote{If $h_k^i[n]$ is a complex Gaussian variable, the bounded difference in \eqref{bound_diff} can be expressed in terms of the Rician $\kappa$-factor as $s_{k}^i[n] \mathcal{B}\frac{2\kappa +1}{\ln 2(\kappa+1)^2}$.}. In fact, in UAV communication systems,  $\mathcal{E}\{\abs{h_k^i[n]}^2\} \gg \mathrm{Var}\{\abs{h_k^i[n]}^2\}$ holds in general and $\mathrm{Var}\{\abs{h_k^i[n]}^2\}$ is generally small \cite{Hourani14Model,khuwaja2018survey}  as the line-of-sight (LoS) path dominates in air-to-ground channels. Also, we note that $\overline{{E\hspace*{-0.5mm}R}}_k^i[n](\mathbf{p},\mathbf{s},\mathbf{r})$ is an upper bound on the expected data rate ${E\hspace*{-0.5mm}R}_k^i[n](\mathbf{p},\mathbf{s},\mathbf{r})$ for the future time slots $n_0 < n \le N_{\mathrm{T}}$. Nevertheless, the data rate of current time slot $n_0$, ${R}_k^i[n_0](\mathbf{p},\mathbf{s},\mathbf{r})$, is known exactly, cf. \eqref{current-rate_k} . Therefore, to facilitate the design of a computationally efficient resource allocation algorithm, we adopt $\overline{{E\hspace*{-0.5mm}R}}_k^i[n](\mathbf{p},\mathbf{s},\mathbf{r})$ as the expected data rate in future time slots in the sequel\footnote{The tightness of the upper bound on the expected data rate will be verified by simulations in Section V.}.
%%%%%%%%%%

\vspace*{-2mm}

\subsection{Optimization Problem Formulation}
At the beginning of each of the $N_{\mathrm{T}}$ time slots, online trajectory design and resource allocation are performed and the real-time trajectory and resource allocation policy are updated accordingly. In particular, in every time slot, we maximize the sum of the achievable throughput in the current time slot and the expected throughput in future time slots. Specifically, in the current time slot $n_0$, we obtain the updated trajectory and resource allocation policy by solving the following problem:
\begin{eqnarray}
\label{online-prob}
&&\hspace*{-15mm} \underset{\mathbf{p},\mathbf{s},\mathbf{r},\mathbf{v},\mathbf{q}}{\maxo} \,\, \frac{1}{N_{\mathrm{F}}\mathcal{B}}
\overset{N_{\mathrm{F}}}{\underset{i = 1}{\sum}} \overset{K}{\underset{k = 1}{\sum}}  \Bigg[ \hspace*{-0.7mm} s_{k}^i[n_0] \mathcal{B} \log_2 \Big( 1 \hspace*{-0.8mm} + \hspace*{-0.8mm} \frac{H_k^i[n_0] p_k^i[n_0]}{\norm{\mathbf{r}[n_0]-\mathbf{r}_k}^2} \Big) \hspace*{-1mm} + \hspace*{-2mm}\overset{N_{\mathrm{T}}}{\underset{n = n_0+1}{\sum}} \hspace*{-1mm} s_{k}^i[n] \mathcal{B} \log_2 \Big( 1 \hspace*{-0.8mm} + \hspace*{-0.8mm} \frac{\mathcal{E}\{H_k^i[n]\} p_k^i[n]}{\norm{\mathbf{r}[n]-\mathbf{r}_k}^2} \hspace*{-0.7mm} \Big) \hspace*{-0.7mm} \Bigg] \notag \\
%%%%%%%%%%%%%%%%
\mbox{s.t.}\hspace*{1mm}
&&\hspace*{-7mm}   \mbox{C1--C13,} \quad \mbox{C14}\mbox{a: } \overset{N_{\mathrm{F}}}{\underset{i = 1}{\sum}} s_{k}^i[n_0] \mathcal{B} \log_2 \Big( 1 + \frac{H_k^i[n_0] p_k^i[n_0]}{\norm{\mathbf{r}[n_0]-\mathbf{r}_k}^2} \Big) \ge R^{\mathrm{req}}_k, \forall k, \notag \\
%%%%%%%%%%
&&\hspace*{-7mm}   \mbox{C14}\mbox{b: } \overset{N_{\mathrm{F}}}{\underset{i = 1}{\sum}} s_{k}^i[n] \mathcal{B} \log_2 \Big( 1 + \frac{\mathcal{E}\{H_k^i[n]\} p_k^i[n]}{\norm{\mathbf{r}[n]-\mathbf{r}_k}^2} \Big)  \ge R^{\mathrm{req}}_k, \forall k, n\in\{n_0 \hspace*{-1mm} + \hspace*{-1mm} 1,\ldots,N_{\mathrm{T}}\}.
\end{eqnarray}
In \eqref{online-prob}, constraints C1--C13 are the same as in the case of offline algorithm design in \eqref{prob} and $\mbox{C14}\mbox{a}$ and $\mbox{C14}\mbox{b}$ specify the minimum required data rate and the expected data rate for user $k$ in the current time slot $n_0$ and the future time slots, respectively. In particular, for the future time slots, since the CSI of the users is not available, $\mbox{C14}\mbox{b}$ imposes a minimum requirement on the upper bound on the expected data rate of the users in any future time slot $n$, $n_0 < n \le N_{\mathrm{T}}$, and the actual expected data rate for user $k$ might be slightly lower than the minimum requirement $R^{\mathrm{req}}_k$.
Nevertheless, in the current time slot, the minimum required data rates of all users are guaranteed since the CSI is known.
%%%%%%
Similar to the problem formulation for the offline algorithm design in \eqref{prob}, problem  \eqref{online-prob} is a mixed-integer non-convex optimization problem which is very difficult to solve. Nevertheless, in the next section, we will develop optimal and suboptimal solutions to problem \eqref{online-prob}.

\vspace*{-5mm}
\subsection{Optimal Solution}
Following the same logic as for solving the offline case and reusing the corresponding variables, we can rewrite \eqref{online-prob} as: \vspace*{-3mm}
\begin{eqnarray}\label{online-equiv-prob}
&&\hspace*{-10mm} \underset{\tilde{\mathbf{p}},\mathbf{r},\mathbf{v},\overline{\mathbf{v}},\mathbf{q},\mathbf{t},\bm{\theta}}{\maxo} \,\, \,\,
\overset{N_{\mathrm{T}}}{\underset{n = n_0}{\sum}} \overset{N_{\mathrm{F}}}{\underset{i = 1}{\sum}} \overset{K}{\underset{k = 1}{\sum}} \log_2 \Big( 1 + \frac{\overline{H}_k^i[n] \tilde{p}_k^i[n] }{\xi \sum_{j\neq k}^{K} \overline{H}_k^i[n] \tilde{p}_j^i[n] + \theta_k[n]} \Big)   \\
%%%%%%
\hspace*{-1mm}\notag\mbox{s.t.}\hspace*{1mm}
&&\hspace*{-7mm}  \overline{\mbox{C1}}\mbox{a}, \mbox{C1b}, \overline{\mbox{C2}}\mbox{a}, \overline{\mbox{C2}}\mbox{b}, \mbox{C3--C11},  \mbox{C15--C18},  \notag \\
%&&\hspace*{-7mm}   \overline{\mbox{C14}}\mbox{a: } \overset{N_{\mathrm{F}}}{\underset{i = 1}{\sum}} \log_2 \Big( 1 + \frac{H_k^i[n_0] \tilde{p}_k^i[n_0]}{\sum_{j\neq k}^{K} H_k^i[n_0]\tilde{p}_j^i[n_0] + \theta_k[n_0]} \Big) \ge R^{\mathrm{req}}_k, \forall k, \notag \\
%%%%%%%
&&\hspace*{-7mm}   \overline{\mbox{C14}}\mbox{: } \overset{N_{\mathrm{F}}}{\underset{i = 1}{\sum}} \mathcal{B} \log_2 \Big( 1 + \frac{\overline{H}_k^i[n] \tilde{p}_k^i[n]}{\sum_{j\neq k}^{K} \overline{H}_k^i[n] \tilde{p}_j^i[n] + \theta_k[n]} \Big)  \ge R^{\mathrm{req}}_k, \forall k, n\in\{n_0,\ldots,N_{\mathrm{T}}\}, \notag
\end{eqnarray}
where $\overline{H}_k^i[n] = H_k^i[n]$ for $n=n_0$ and $\overline{H}_k^i[n] = \mathcal{E}\{H_k^i[n]\}$ for $n_0+1 \le n \le N_{\mathrm{T}}$.
Then, for facilitating monotonic optimization, we impose the following constraint: \vspace*{-3mm}
\begin{eqnarray}
&&\hspace*{-7mm}   \overline{\mbox{C19}}\mbox{: } 1 \le \chi_{k}^{i}[n] \le \frac{\overline{f}_{k}^{i}[n](\tilde{\mathbf{p}},\bm{\theta})}{\overline{g}_{k}^{i}[n](\tilde{\mathbf{p}},\bm{\theta})},
\end{eqnarray}
where
$\overline{f}_{k}^{i}[n](\tilde{\mathbf{p}},\bm{\theta})= \overline{H}_k^i[n]\tilde{p}_k^i[n] + \xi \sum_{j \neq k}^{K} \overline{H}_k^i[n]\tilde{p}_j^i[n] + \theta_k[n]$ and $\overline{g}_{k}^{i}[n](\tilde{\mathbf{p}},\bm{\theta})\hspace*{-0mm}=
\hspace*{-1mm} \xi \sum_{j \neq k}^{K} \overline{H}_k^i[n]\tilde{p}_j^i[n] + \theta_k[n]$.
Now, the problem in \eqref{online-equiv-prob} can be rewritten in the form of a standard monotonic optimization problem as: \vspace*{-4mm}
\begin{eqnarray}\label{online-MO-pro}
\hspace*{-1mm} \underset{\bm{\chi},\bm{\mu},\bm{\varpi},\bm{\tau}}{\maxo}\hspace*{-1mm} && \hspace*{-1mm} \overset{N_{\mathrm{T}}}{\underset{n = n_0}{\sum}} \overset{N_{\mathrm{F}}}{\underset{i = 1}{\sum}} \overset{K}{\underset{k = 1}{\sum}} \log_2(\chi_{k}^{i}[n])  \quad \quad \mbox{s.t.}\hspace*{1mm}
 (\bm{\chi},\bm{\mu},\bm{\varpi},\bm{\tau})\in\mathcal{V}=\mathcal{G} \cap \mathcal{H},
\end{eqnarray}
where feasible set $\mathcal{G}$ is spanned by constraints $\overline{\mbox{C1}}\mbox{a}$, $\mbox{C1b}$, $\overline{\mbox{C2}}\mbox{a}$, $\overline{\mbox{C2}}\mbox{b}$, $\mbox{C3--C11}$, $\mbox{C15}$, and $\mbox{C17}\mbox{--}\overline{\mbox{C19}}$ and feasible set $\mathcal{H}$ is spanned by constraints $\overline{\mbox{C14}}$ and C16. The optimal solution of the monotonic optimization problem in \eqref{online-MO-pro} can be obtained by applying the sequential polyblock approximation algorithm summarized in \textbf{Algorithm 1}.

The proposed monotonic optimization based algorithm finds the globally optimal online trajectory and resource allocation policy\footnote{The obtained optimal online resource allocation policy is optimal in the sense that it maximizes the sum of the achievable rate for the current time slot and the upper bound on the expected rate for future time slots.}.
However, the computational complexity of the algorithm grows exponentially with the number of time slots and users which is prohibitive for real-time operation of UAV-based communication systems. Hence, in order to reduce complexity, in the next section, we develop a computationally efficient suboptimal scheme which finds a locally optimal policy with polynomial time complexity. Nevertheless, the optimal online scheme is a useful benchmark scheme as it provides a quantitative basis for comparison for any suboptimal algorithm.

\vspace*{-2mm}
\subsection{Suboptimal Solution}
Since problem \eqref{online-equiv-prob} is equivalent to the original online resource allocation problem in \eqref{online-prob}, we focus on the solution of \eqref{online-equiv-prob} in this section.
With the equivalent constraints introduced in Section \ref{Offline_optimal_section}, we can rewrite problem \eqref{online-equiv-prob} as: \vspace*{-2mm}
\begin{eqnarray}\label{subopt-prob}
&&\hspace*{-1mm} \underset{\tilde{\mathbf{p}},\mathbf{r},\mathbf{v},\overline{\mathbf{v}},\mathbf{q},\mathbf{t},\bm{\theta},\bm{\mu}}{\maxo} \,\, \,\, \overset{N_{\mathrm{T}}}{\underset{n = n_0}{\sum}} \overset{N_{\mathrm{F}}}{\underset{i = 1}{\sum}} \overset{K}{\underset{k = 1}{\sum}} \log_2 \Big( 1 + \frac{\overline{H}_k^i[n] \tilde{p}_k^i[n] }{\xi \sum_{j\neq k}^{K} \overline{H}_k^i[n]\tilde{p}_j^i[n] + \theta_k[n]} \Big)   \\[-2mm]
\hspace*{-1mm}\notag\mbox{s.t.}\hspace*{1mm}
&&\hspace*{-7mm}  \overline{\mbox{C1}}\mbox{a}, \mbox{C1b}, \mbox{C3--C11}, \mbox{C15}, \mbox{C17}, \quad  \quad \overline{\mbox{C2}}\mbox{: } \ln\big(\varpi[n]\big) - \ln\Big(\frac{C_1\Delta_{\mathrm{T}}}{1+e^{-k_{c}(z[n]-\alpha)}}\Big) \le 0, \,\, \forall n, \notag \\[-3mm]
%%%%%%%%%%%%%%%%%%%%%%%%%%%%%%%
&&\hspace*{-7mm}   \overline{\mbox{C14}}\mbox{: } \overset{N_{\mathrm{F}}}{\underset{i = 1}{\sum}} \mathcal{B} \log_2 \Big( 1 + \frac{\overline{H}_k^i[n] \tilde{p}_k^i[n]}{\sum_{j\neq k}^{K} \overline{H}_k^i[n] \tilde{p}_j^i[n] + \theta_k[n]} \Big)  \ge R^{\mathrm{req}}_k, \forall k, n\in\{n_0,\ldots,N_{\mathrm{T}}\}, \notag \\[-1mm]
&&\hspace*{-7mm} \mbox{C16: }  \mu[n] \ge \frac{1}{\sqrt{\norm{(v_x[n], v_y[n])}^2 \hspace*{-0.5mm} + \hspace*{-0.5mm} \sqrt{\norm{(v_x[n], v_y[n])}^4 \hspace*{-0.5mm} + \hspace*{-0.5mm} 4 V_{\mathrm{h}}^4}}}, \,\,  \forall n, \notag \\[-1mm]
&&\hspace*{-7mm}   \mbox{C18: } 0 \le q[n+1] - q[n] + t[n]\Delta_{\mathrm{T}} -C_2\Delta_{\mathrm{T}} \le \varpi[n]. \notag
\end{eqnarray}
In \eqref{subopt-prob}, constraint C16 is non-convex which is an obstacle for the design of a computationally efficient algorithm. In order to overcome this difficulty, we rewrite C16 in equivalent form as follows: \vspace*{-2mm}
\begin{eqnarray}\label{Equiv-C16}
&&\hspace*{-7mm}  \mbox{C16a: } \mu[n] \ge \frac{1}{b[n]}, \forall n, \hspace*{39.5mm} \mbox{C16b: } \big(b[n]\big)^2 \le l[n] + \sqrt{\gamma[n]},  \forall n,   \\[-2mm]
%%%%
&&\hspace*{-7mm}  \mbox{C16c: } l[n] \le \big(v_x[n]\big)^2 + \big(v_y[n]\big)^2,  \forall n,  \hspace*{15mm} \mbox{C16d: } \gamma[n] \le \big(l[n]\big)^2 + 4 V_{\mathrm{h}}^4,  \forall n, \\[-2mm]
&&\hspace*{-7mm}  \mbox{C16e: } b[n], l[n], \gamma[n] \ge 0, \forall n,
\end{eqnarray}
where $b[n]$, $l[n]$, and $\gamma[n]$ are auxiliary optimization variables. Yet, C16c and C16d are still non-convex constraints.
In addition, in \eqref{subopt-prob}, constraints $\overline{\mbox{C2}}$ and $\overline{\mbox{C14}}$ and the objective function are also non-convex. To overcome this difficulty, we first rewrite \eqref{subopt-prob} in equivalent form as a difference of convex (d.c.) programming problem \cite{ng2015power}:\vspace*{-1mm}
\begin{eqnarray}\label{dc-subopt-prob}
&&\hspace*{-11mm} \underset{\tilde{\mathbf{p}},\mathbf{r},\mathbf{v},\overline{\mathbf{v}},\mathbf{q},\mathbf{t},\bm{\theta},\bm{\mu},\mathbf{b},\mathbf{l},\bm{\gamma}} {\mino} \,\, \,\, -\overset{N_{\mathrm{T}}}{\underset{n = n_0}{\sum}} \overset{N_{\mathrm{F}}}{\underset{i = 1}{\sum}} \overset{K}{\underset{k = 1}{\sum}} \log_2 \Big( \overline{H}_k^i[n]\tilde{p}_k^i[n] + \xi \sum_{j\neq k}^{K} \overline{H}_k^i[n]\tilde{p}_j^i[n] + \theta_k[n] \Big) - G(\tilde{\mathbf{p}},\bm{\theta})  \\[-2mm]
%%%%%%%%%%%%%
\hspace*{-1mm}\notag\mbox{s.t.}\hspace*{1mm}
&&\hspace*{-7mm}  \overline{\mbox{C1}}\mbox{a}, \mbox{C1b}, \mbox{C3--C11}, \mbox{C15}, \mbox{C16a}, \mbox{C16b}, \mbox{C16e}, \mbox{C17}, \mbox{C18}, \notag \\[-2mm]
%%%%
&&\hspace*{-7mm} \overline{\mbox{C2}}\mbox{: } - \ln\Big(\frac{C_1\Delta_{\mathrm{T}}}{1+e^{-k_{c}(z[n]-\alpha)}}\Big) + \ln\big(\varpi[n]\big)  \le 0, \forall n, \notag \\[-2mm]
%%%%%%%%%%%%%%%%%%%%%%%%%%%%%%%
&&\hspace*{-7mm}   \overline{\mbox{C14}}\mbox{: } \hspace*{-1mm} - \hspace*{-0.6mm} \overset{N_{\mathrm{F}}}{\underset{i = 1}{\sum}} \hspace*{-0.4mm} \mathcal{B} \log_2 \hspace*{-0.6mm} \Big( \hspace*{-0.4mm} \sum_{j=1}^{K} \overline{H}_k^i[n]\tilde{p}_j^i[n]\hspace*{-0.6mm}  + \hspace*{-0.6mm} \theta_k[n] \hspace*{-0.4mm} \Big) \hspace*{-0.6mm}  - \hspace*{-0.6mm}Q_k[n](\tilde{\mathbf{p}},\bm{\theta}) \hspace*{-0.6mm} \le  \hspace*{-0.6mm}- R^{\mathrm{req}}_k, \forall k, n \hspace*{-1mm} \in \hspace*{-1mm} \{n_0,\hspace*{-0.6mm}\ldots\hspace*{-0.6mm},N_{\mathrm{T}}\}, \notag \\[-2mm]
%%%%%%%%%%%%%
&&\hspace*{-7mm} \mbox{C16c: } l[n] - \big(v_x[n]\big)^2 - \big(v_y[n]\big)^2 \le 0,  \forall n, \hspace*{15mm} \mbox{C16d: } \gamma[n] - \big(l[n]\big)^2 \le 4 V_{\mathrm{h}}^4, \forall n, \notag
\end{eqnarray}
where \vspace*{-3mm}
\begin{eqnarray}
G(\tilde{\mathbf{p}},\bm{\theta}) \hspace*{-1.5mm} &=& \hspace*{-1.5mm} -\overset{N_{\mathrm{T}}}{\underset{n = n_0}{\sum}} \overset{N_{\mathrm{F}}}{\underset{i = 1}{\sum}} \overset{K}{\underset{k = 1}{\sum}} \log_2 \Big( \xi \sum_{j \neq k}^{K} \overline{H}_k^i[n]\tilde{p}_j^i[n] + \theta_k[n] \Big), \\[-2mm]
%%%%%%%%%%
Q_k[n](\tilde{\mathbf{p}},\bm{\theta}) \hspace*{-1.5mm} &=& \hspace*{-1.5mm} -\overset{N_{\mathrm{F}}}{\underset{i = 1}{\sum}} \mathcal{B}\log_2 \Big( \sum_{j \neq k}^{K} \overline{H}_k^i[n]\tilde{p}_j^i[n] + \theta_k[n] \Big),
\end{eqnarray}
and $\mathbf{b}\in\mathbb{R}^{N_{\mathrm{T}}\times 1}$, $\mathbf{l}\in\mathbb{R}^{N_{\mathrm{T}}\times 1}$, and $\bm{\gamma}\in\mathbb{R}^{N_{\mathrm{T}}\times 1}$ are the collections of all $b[n]$, $l[n]$, and $\gamma[n]$, respectively.
We note that the objective function and the constraint functions in $\overline{\mbox{C2}}$, $\overline{\mbox{C14}}$, C16c, and C16d are differences of convex  functions while the other constraints are convex. Locally optimal solutions for d.c. programming problems, e.g. \eqref{dc-subopt-prob}, can be obtained by applying successive convex approximation \cite{dinh2010local}. In particular, for any point $\tilde{\mathbf{p}}^{(m)}$ and $\bm{\theta}^{(m)}$, we have \vspace*{-3mm}
\begin{eqnarray}\label{ineq1}
G(\tilde{\mathbf{p}},\bm{\theta}) &\ge& G(\tilde{\mathbf{p}}^{(m)},\bm{\theta}^{(m)}) + \nabla_{\tilde{\mathbf{p}}} G(\tilde{\mathbf{p}}^{(m)},\bm{\theta}^{(m)})^T(\tilde{\mathbf{p}}-\tilde{\mathbf{p}}^{(m)})
\hspace*{-0.7mm}+\hspace*{-0.7mm} \nabla_{\bm{\theta}} G(\tilde{\mathbf{p}}^{(m)},\bm{\theta}^{(m)})^T(\bm{\theta}-\bm{\theta}^{(k)}) \notag \\
&=& G(\tilde{\mathbf{p}}^{(m)},\bm{\theta}^{(m)}) -\overset{N_{\mathrm{T}}}{\underset{n = n_0}{\sum}} \overset{N_{\mathrm{F}}}{\underset{i = 1}{\sum}} \overset{K}{\underset{k = 1}{\sum}} \frac{\xi \overline{H}_k^i[n]\big(\tilde{p}_j^i[n]-(\tilde{p}_j^i)^{(m)}[n]\big) + \theta_k[n]-\theta_k^{(m)}[n]}{ \big(\xi \sum_{j \neq k}^{K} \overline{H}_k^i[n](\tilde{p}_j^i)^{(m)}[n] + \theta_k^{(m)}[n]\big) \ln 2} \notag \\
&\triangleq& \underline{G}(\tilde{\mathbf{p}},\bm{\theta},\tilde{\mathbf{p}}^{(m)},\bm{\theta}^{(m)}),
\end{eqnarray}
where the right hand side of \eqref{ineq1} is the summation of affine functions representing a global underestimator of $G(\tilde{\mathbf{p}},\bm{\theta})$.
Similarly, we can construct global underestimators for the subtrahend convex functions in $\overline{\mbox{C14}}$, C16c, and C16d.
%%%%%%%%%
\begin{table} \vspace*{-8mm}
%\caption{Iterative Resource Allocation Algorithm for Suboptimal Online and Optimal Offline Designs.}\label{table:algorithm}\vspace*{-0cm}
%\small
\begin{algorithm} [H]                    % enter the algorithm environment
\caption{Successive Convex Approximation}    \vspace*{-0.2mm}      % give the algorithm a caption
\label{alg3}                           % and a label for \ref{} commands later in the document
\begin{algorithmic} [1]
\small          % enter the algorithmic environment
\STATE Initialize iteration index $m=1$ and initial point $\bm{\Psi}^{(1)}$ and set error tolerance $\epsilon_3 \ll 1$ \vspace*{-0.5mm}

\REPEAT \vspace*{-0.2mm}
%%%%%%%%%%%%%%%%%%%%
\STATE For given $\bm{\Psi}^{(m)}$, solve the convex problem in \eqref{sca-subopt-prob} and store the intermediate solution $\bm{\Xi}$ and $\bm{\Psi}$

\STATE Set $m=m+1$ and $\bm{\Psi}^{(m)}=\bm{\Psi}$ \vspace*{-0.2mm}

\UNTIL $\frac{\norm{\bm{\Psi}^{(m)} -\bm{\Psi}^{(m-1)}}} {\norm{\bm{\Psi}^{(m-1)}}} \le \epsilon_3$
 \vspace*{-0.2mm}

\STATE Obtain final resource allocation policy $\bm{\Xi}^{*}=\bm{\Xi}^{(m)}$, $\bm{\Psi}^{*}=\bm{\Psi}^{(m)}$
\end{algorithmic}
\end{algorithm}\vspace*{-14mm}
\end{table}
%%%%%%%%%%%%%%%%%%%%%
Besides, we define $\bm{\Psi}$, $\bm{\Psi}^{(m)}$, $\bm{\Xi}$, and $\bm{\Xi}^{(m)}$ as the collection of variables
$\{\tilde{\mathbf{p}}, \hspace*{-0.5mm}\mathbf{v}, \hspace*{-0.5mm}\bm{\varpi}, \hspace*{-0.5mm}\bm{\theta}, \hspace*{-0.5mm} \mathbf{l}\}$,
$\{\tilde{\mathbf{p}}^{(m)},\hspace*{-0.5mm}\mathbf{v}^{(m)}, \hspace*{-0.5mm}\bm{\varpi}^{(m)}, \hspace*{-0.5mm}\bm{\theta}^{(m)}, \hspace*{-0.5mm}\mathbf{l}^{(m)}\}$,
$\{\mathbf{r}, \hspace*{-0.5mm}\mathbf{q}, \hspace*{-0.5mm}\mathbf{t}, \hspace*{-0.5mm}\overline{\mathbf{v}},\hspace*{-0.5mm}\bm{\mu}, \hspace*{-0.5mm}\mathbf{b}, \hspace*{-0.5mm}\bm{\gamma}\}$, and $\{\mathbf{r}^{(m)}, \hspace*{-0.5mm}\mathbf{q}^{(m)}, \hspace*{-0.5mm}\mathbf{t}^{(m)}, \hspace*{-0.5mm}\overline{\mathbf{v}}^{(m)}, \hspace*{-0.5mm}\bm{\mu}^{(m)}, \hspace*{-0.5mm}\mathbf{b}^{(m)}, \hspace*{-0.5mm}\bm{\gamma}^{(m)}\}$,  respectively.
%%%%%%%%%%%%%%%%%%%
Then, for any given $\bm{\Psi}^{(m)}$, we can find a lower bound of \eqref{dc-subopt-prob} by solving the following optimization problem: \vspace*{-1mm}
\begin{eqnarray}\label{sca-subopt-prob}
&&\hspace*{-8mm} \underset{\bm{\Xi},\bm{\Psi}} {\mino} \,\, \,\, -\overset{N_{\mathrm{T}}}{\underset{n = n_0}{\sum}} \overset{N_{\mathrm{F}}}{\underset{i = 1}{\sum}} \overset{K}{\underset{k = 1}{\sum}} \log_2 \Big( \overline{H}_k^i[n]\tilde{p}_k^i[n] + \xi \sum_{j \neq k}^{K} \overline{H}_k^i[n]\tilde{p}_j^i[n] + \theta_k[n] \Big) - \underline{G}(\tilde{\mathbf{p}},\bm{\theta},\tilde{\mathbf{p}}^{(m)},\bm{\theta}^{(m)})  \notag \\[-2mm]
%%%%%%%%%%%%%
\hspace*{-1mm}\notag\mbox{s.t.}\hspace*{1mm}
&&\hspace*{-7mm}  \overline{\mbox{C1}}\mbox{a}, \mbox{C1b}, \mbox{C3--C11}, \mbox{C15}, \mbox{C16a}, \mbox{C16b}, \mbox{C16e}, \mbox{C17}, \mbox{C18},\notag \\[-2mm]
%%%%
&&\hspace*{-7mm}   \widetilde{\mbox{C2}}\mbox{: } - \ln\Big(\frac{C_1\Delta_{\mathrm{T}}}{1+e^{-k_{c}(z[n]-\alpha)}}\Big) + \ln\big(\varpi^{(m)}[n]\big) + \frac{\varpi[n]-\varpi^{(m)}[n]}{\varpi^{(m)}[n]} \le 0, \forall n, \notag \\[-2mm]
%%%%%%%%%%%%%%%%%%%%%%%%%%%%%%%
&&\hspace*{-7mm}   \widetilde{\mbox{C14}}\mbox{: } -\overset{N_{\mathrm{F}}}{\underset{i = 1}{\sum}} \mathcal{B} \log_2 \hspace*{-0.5mm} \Big( \hspace*{-0.5mm} \sum_{j=1}^{K} H_k^i[n]\tilde{p}_j^i[n] \hspace*{-0.5mm} + \hspace*{-0.5mm} \theta_k[n] \hspace*{-0.5mm} \Big) \hspace*{-0.5mm} - \hspace*{-0.5mm} \underline{Q}_k[n](\tilde{\mathbf{p}},\bm{\theta},\tilde{\mathbf{p}}^{(m)},\bm{\theta}^{(m)}) \hspace*{-0.5mm} \le \hspace*{-0.5mm} - \hspace*{-0.5mm} R^{\mathrm{req}}_k, \notag \\[-2mm]
%\forall k, n  \in \{n_0,\ldots,N_{\mathrm{T}}\}, \notag \\[-2mm]
%%%%%%%%%%%%%
&&\hspace*{-7mm} \widetilde{\mbox{C16c}}\mbox{: } l[n] - 2 v_x^{(m)}[n]v_x[n] - 2 v_y^{(m)}[n]v_y[n] + \big(v_x^{(m)}[n]\big)^2 + \big(v_x^{(m)}[n]\big)^2  \le 0,  \forall n, \notag \\[-2mm]
&&\hspace*{-7mm} \widetilde{\mbox{C16d}}\mbox{: } \gamma[n] - 2 l^{(m)}[n] l[n]  + \big(l^{(m)}[n]\big)^2 \le 4 V_{\mathrm{h}}^4, \forall n,
\end{eqnarray}
where \vspace*{-3mm}
\begin{eqnarray}
\underline{Q}_k[n](\tilde{\mathbf{p}},\hspace*{-0.8mm}\bm{\theta},\hspace*{-0.8mm}\tilde{\mathbf{p}}^{(m)},\hspace*{-0.8mm}\bm{\theta}^{(m)})  \hspace*{-0.8mm}= \hspace*{-0.8mm}Q_k[n](\tilde{\mathbf{p}}^{(m)},\hspace*{-0.8mm}\bm{\theta}^{(m)})- \mathcal{B}\overset{N_{\mathrm{F}}}{\underset{i = 1}{\sum}} \frac{H_k^i[n]\big(\tilde{p}_j^i[n] \hspace*{-0.8mm} - \hspace*{-0.8mm} (\tilde{p}_j^i)^{(m)}[n]\big) \hspace*{-0.8mm} + \hspace*{-0.8mm} \theta_k[n]-\theta_k^{(m)}[n]}{ \big(\sum_{j \neq k}^{K} H_k^i[n](\tilde{p}_j^i)^{(m)}[n] \hspace*{-0.8mm} + \hspace*{-0.8mm} \theta_k^{(m)}[n]\big) \ln 2}
\end{eqnarray}
represents a global underestimator for $Q_k[n](\tilde{\mathbf{p}},\bm{\theta})$.
%%%%
We note that the problem in \eqref{sca-subopt-prob} is convex and can be solved by standard optimization problem solvers such as CVX \cite{website:CVX}. Then, we can tighten the obtained lower bound by applying the iterative algorithm summarized in \textbf{Algorithm 3}. In particular, by solving the lower bound problem in \eqref{sca-subopt-prob} in each iteration, the proposed iterative scheme generates a sequence of solutions $\tilde{\mathbf{p}}^{(m+1)}$, $\mathbf{r}^{(m+1)}$, $\mathbf{v}^{(m+1)}$, and $\mathbf{q}^{(m+1)}$ successively. It can be shown that the proposed suboptimal algorithm converges to a locally optimal solution of \eqref{dc-subopt-prob} and has polynomial time computational complexity \cite{dinh2010local}.

\begin{table}[t]\vspace*{-4mm}\caption{System parameters}\vspace*{-4mm}\label{tab:parameters}
\newcommand{\tabincell}[2]{\begin{tabular}{@{}#1@{}}#2\end{tabular}}\vspace*{-0mm}
\centering
\begin{tabular}{|l|l|}\hline
\hspace*{-1mm}Carrier center frequency and bandwidth & $700$ $\mathrm{MHz}$ and $5$ $\mathrm{MHz}$ \cite{Tech:LTEUAVQualcomm} \\
\hline
\hspace*{-1mm}Number and bandwidth of subcarriers & $64$  and $78$ $\mathrm{kHz}$\\
\hline
%\hspace*{-1mm}Parameters for atmospheric transmittance, $\alpha$, $\beta$   & $0.8978$, $0.2804$ \cite{duffie2013solar} \\
%\hline
\hspace*{-1mm}Average solar radiation and efficiency of solar panels,  $G$ and $\eta$ & $1367$ $\mathrm{W}\hspace*{-0.4mm}/\hspace*{-0.4mm}\mathrm{m}^2$ and $0.4$ \\
\hline
\hspace*{-1mm}Altitude of cloud, $L_{\mathrm{low}}$ and $L_{\mathrm{up}}$  & $700$ $\mathrm{m}$ and $1400$ $\mathrm{m}$ \cite{kokhanovsky2004optical} \\
\hline
\hspace*{-1mm}Absorption coefficient of cloud and area of the solar panel, $\beta_c$ and $S$ &  $0.01$ \cite{kokhanovsky2004optical} and $1 \mathrm{m}^2$\\
\hline
\hspace*{-1mm}Parameters of the lower bound on the harvest solar energy, $k_c$ and $\alpha$ &  $0.05$ and $1351$\\
\hline
\hspace*{-1mm}Altitude limitation for UAV, $z_{\mathrm{min}}$ and $z_{\mathrm{max}}$ & $100$ $\mathrm{m}$ and $1600$ $\mathrm{m}$  \\
\hline
\hspace*{-1mm}Duration of each time slot and maximum transmit power of the UAV, $\Delta_{\mathrm{T}}$ and $P_{\mathrm{max}}$  & 0.02 $\mathrm{s}$ and $42$  $\mathrm{dBm}$ \cite{Tech:LTEUAVQualcomm}\\
\hline
\hspace*{-1mm}RF power amplifier efficiency and noise power spectral density, $\varepsilon$ and $\mathcal{N}_0$   &  $0.5$ and $-174$ $\mathrm{dBm/Hz}$  \\
\hline
\hspace*{-1mm}Maximum horizontal and vertical speed of UAV, $V_{\mathrm{max}}^{\mathrm{xy}}$ and  $V_{\mathrm{max}}^{\mathrm{z}}$&  $10$ $\mathrm{m/s}$ and $4$ $\mathrm{m/s}$   \\
\hline
\hspace*{-1mm}Total area of rotor disks and density of air, $A$ and $\rho$ & $0.18$ $\mathrm{m}^2$ and $1.225 \mathrm{kg/m^3}$ \\
\hline
\hspace*{-1mm}Mass of the UAV and the gravitational acceleration, $m$ and $g$ & $4$ $\mathrm{kg}$ and $9.8$ $\mathrm{m/s^2}$ \cite{oettershagen2016perpetual} \\
\hline
\hspace*{-1mm}Profile drag coefficient $C_{\mathrm{D0}}$ &   $0.08$ \cite{seddon2011basic} \\
\hline
\hspace*{-1mm}Static power consumption and maximum acceleration of UAV, $P_{\mathrm{static}}$ and $a_{\mathrm{max}}$ &   $5$ $\mathrm{W}$ and $2 \mathrm{m/s}^2$ \cite{oettershagen2016perpetual,seddon2011basic}\\
\hline
\hspace*{-1mm}Initial and minimum required remaining stored energy, $q_{0}$ and $q_{\mathrm{end}}$   &  $111$ $\mathrm{Wh}$ and $55$ $\mathrm{Wh}$ \cite{oettershagen2016perpetual}\\
\hline
\hspace*{-1mm}Capacity of the on-board battery, $q_{\mathrm{max}}$    &  $222$ $\mathrm{Wh}$ \cite{suarez2017lightweight} \\
\hline
\hspace*{-1mm}Error tolerances $\epsilon_1$, $\epsilon_2$, and $\epsilon_3$ for {\bf{Algorithms 1}}, {\bf{2}}, and  {\bf{3}} &  $0.01$  \\
\hline
\hspace*{-1mm}Penalty factor, $\xi$ &  $1 \times 10^{20}$  \\
\hline
\end{tabular}
\vspace*{-6mm}
\end{table}

\vspace*{-0mm}
\section{Simulation Results}
In this section, we evaluate the system performance of the proposed schemes via simulations. The adopted simulation parameters are given in Table \ref{tab:parameters}.
We consider a single circular cell where the $K$ downlink users are randomly and uniformly distributed within the cell with radius $800$ meter.
The initial point of the UAV's trajectory is set at the origin of the cell and the minimum allowable altitude, i.e., $\mathbf{r}[0] = (0,0,z_{\mathrm{min}})$.
In this work, we assume that the entire service area is covered by clouds and we consider the trajectory and resource allocation design during daytime.
In each time slot, the small-scale fading coefficients of the channels between the UAV and the downlink users on each subcarrier are independent and identically distributed random variables following a Rician distribution with Rician factor $6$ dB.
We assume that the minimum QoS requirements for the users is $R_{k}^{\mathrm{req}}=\frac{R_{\mathrm{req}}}{K}$, where $R_{\mathrm{req}}=50$ Mbits/s.
% We obtained the results shown in this section by averaging over $5000$ realizations of multipath fading.

Besides, we also consider the performance of two offline baseline schemes for comparison.
For baseline scheme 1, we assume the availability of perfect CSI knowledge and the horizontal position of the UAV is fixed at the origin of the cell, i.e., $(x,y)=(0,0)$. Then, we jointly optimize the flight altitude $z$, velocity, transmit power, and subcarrier allocation.
For baseline scheme 2, we assume that the UAV first climbs up to altitude $z=\min\{L_{\mathrm{up}},z_{\mathrm{max}}\}$ and then stays at that altitude. Besides, the user on each subcarrier is selected at random. Then, we optimize the $(x,y)$ coordinates and the transmit power in an offline manner assuming perfect CSI knowledge.

\begin{figure}[t]
\centering\vspace*{-5mm}
\begin{minipage}[b]{0.45\linewidth} \hspace*{-1cm}
    \includegraphics[width=3.55in]{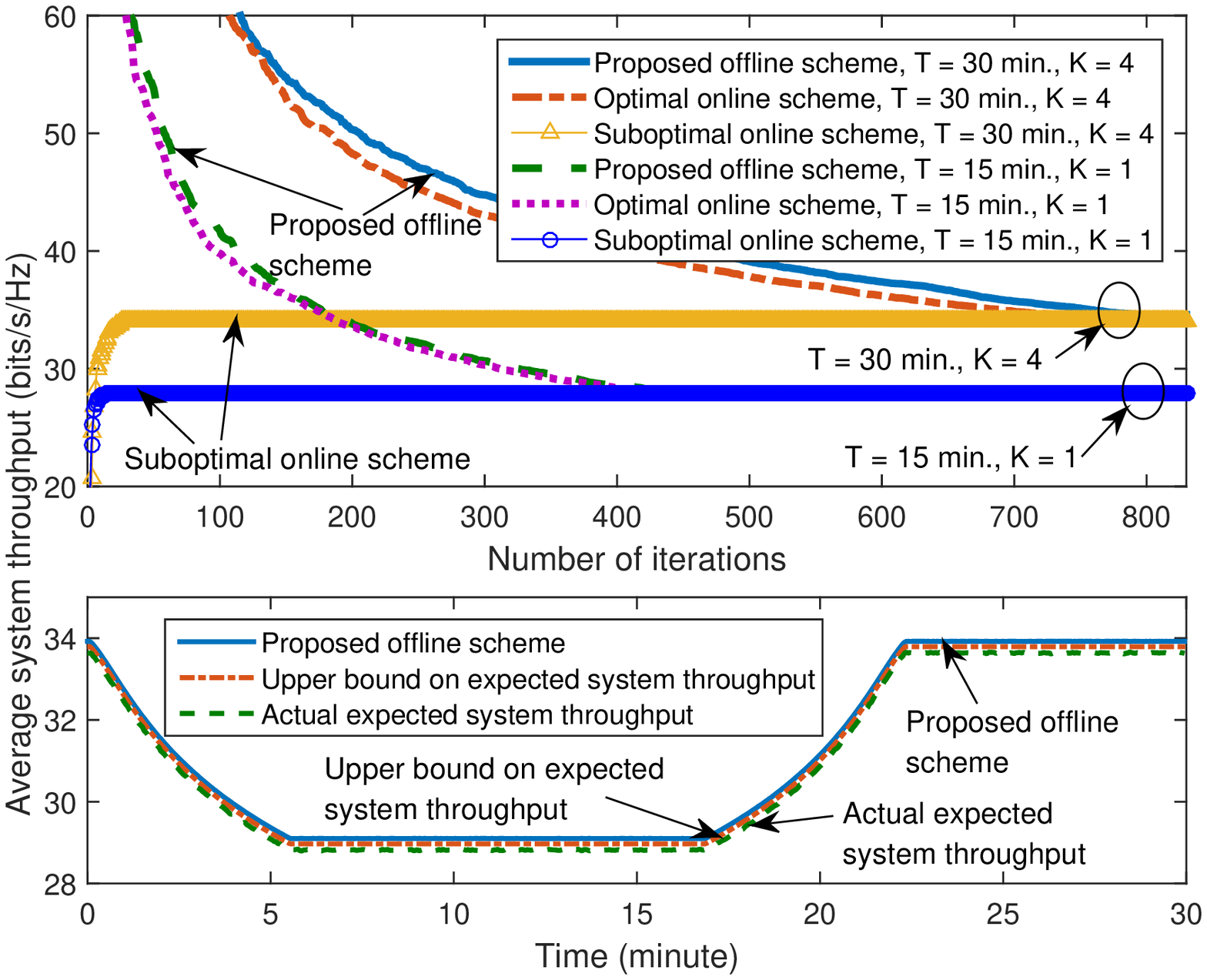}\vspace*{-7mm}
\caption{Top figure: convergence of different resource allocation schemes; Bottom figure: expected system throughput for different schemes.}
\label{fig:convergence_vs_iteration}
\end{minipage}\hspace*{8mm}
\begin{minipage}[b]{0.45\linewidth} \hspace*{-1cm}
    \includegraphics[width=3.55in]{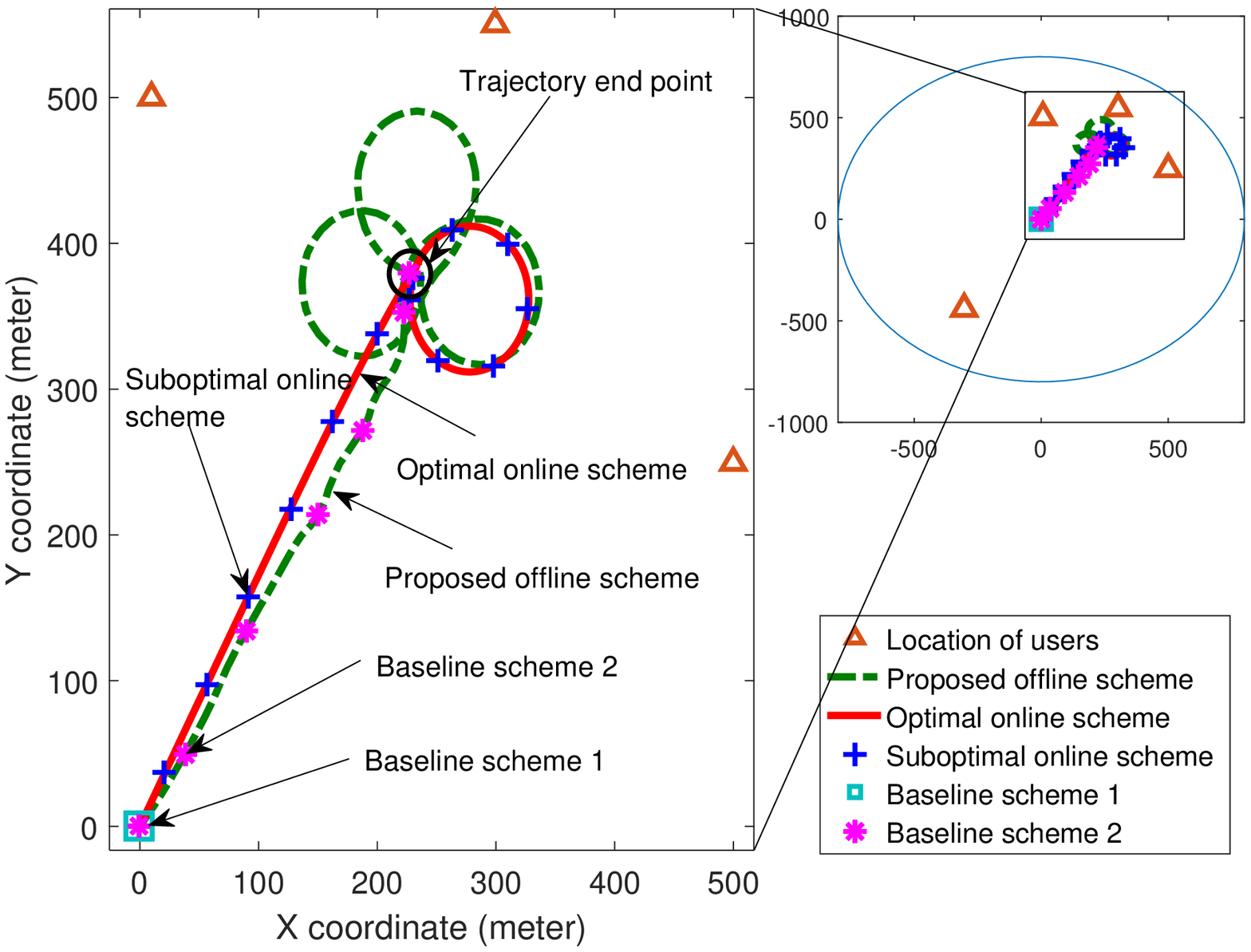}\vspace*{-7mm}
\caption{Trajectory in the horizontal plane for $T = 15$ minutes, $K = 4$, and different resource allocation schemes.}
\label{fig:trajectory_xy}
\end{minipage}\vspace*{-5mm}
\end{figure}

\vspace*{-2mm}
\subsection{Convergence of Proposed Algorithms and Accuracy of Upper Bound on Expected Data Rate}
In Figure \ref{fig:convergence_vs_iteration} (top figure), we investigate the convergence of the proposed offline scheme and the optimal and suboptimal online schemes, for different time durations $T$ and different numbers of users $K$. For a fair comparison, for the online scheme, we focus on the convergence of the algorithms for the first time slot, i.e., $n_0 = 1$.
As can be seen, the speed of convergence of the proposed offline scheme is similar to that of the optimal online scheme as both schemes are based on monotonic optimization.
Besides, Figure \ref{fig:convergence_vs_iteration} also reveals that the proposed optimal and suboptimal online schemes converge to the optimal offline resource allocation solution for all considered values of $T$ and $K$, while the suboptimal online scheme requires substantially fewer iterations to converge. % In addition, the converged optimal online solution is close to the performance obtained by the proposed benchmark offline scheme.
In particular, for $T = 15$ minutes and $K=1$, the proposed optimal and suboptimal online schemes converge to the optimal online solution in less than $430$ and $20$ iterations, respectively. For the case of $T = 30$ minutes and $K = 4$, the proposed optimal offline and online schemes need considerably more iterations to converge due to the exponentially enlarged search space spanned by the larger numbers of users and time slots. In contrast, as can be seen from Figure \ref{fig:convergence_vs_iteration}, the number of iterations required for the proposed suboptimal online scheme to converge is less sensitive to the length of the transmission period and the number of users, which demonstrates its practicality.
%%%%%%
On the other hand, in Figure \ref{fig:convergence_vs_iteration} (bottom figure), we also show the expected system throughput over the period of $1 < n \le N_{\mathrm{T}}$ for $T = 30$ minutes, $K = 4$ users, and $n_0 = 1$. The curve of the upper bound on the expected system throughput is obtained by solving problem \eqref{online-MO-pro} and the resulting resource allocation policy is defined as $\bm{\Pi}$.
%%%%%%%%
Besides, we calculate the actual expected system throughput by substituting $\bm{\Pi}$ into $\sum_{i=1}^{N_{\mathrm{F}}} \sum_{k=1}^{K} \frac{{E\hspace*{-0.5mm}R}_k^i[n](\mathbf{p},\mathbf{s},\mathbf{r})}{N_\mathrm{F} \mathcal{B}}$, where ${E\hspace*{-0.5mm}R}_k^i[n](\mathbf{p},\mathbf{s},\mathbf{r})=\frac{s_{k}^i[n] \mathcal{B}}{M} \sum_{f=1}^{M}\log_2 \Big( 1 + \frac{\tilde{H}_{k,f}^i[n] p_k^i[n]}{\norm{\mathbf{r}[n]-\mathbf{r}_k}^2} \Big)$, $\tilde{H}_{k,f}^i[n]$ are random channel realizations, and $M=10000$. As can be observed, the performance of the actual expected system throughput and the upper bound on the expected system throughput closely approach the performance of the proposed optimal offline scheme. This indicates that the adopted upper bound $\overline{{E\hspace*{-0.5mm}R}}_k^i[n](\mathbf{p},\mathbf{s},\mathbf{r})$ is a good approximation for the actual expected data rate of UAV communication systems for the adopted typical system parameters.

\vspace*{-2mm}
\subsection{Trajectory}
Figure \ref{fig:trajectory_xy} depicts the projection of the trajectory of the UAV onto the horizontal X-Y plane for $T = 15$ minutes, $K = 4$ user, and different resource allocation schemes.
As can be observed, at the beginning, the proposed offline and online schemes and baseline scheme 2 yield similar aerial trajectories which have the UAV first head towards the center of the locations of the users in order to improve the system sum throughput. Subsequently, for the offline and online scheme, the UAV cruises around near the centroid of a virtual triangle connecting the three users that are near each other, since according to \eqref{aero_power_consumption}, the UAV consumes less aerodynamic power during level flight compared to hovering.
On the other hand, the UAV in baseline scheme 2 hovers in the air after arriving at the centroid of the three users that are near each other. This is because, with baseline scheme 2, the UAV is required to fly above the cloud to obtain sufficient solar energy supply.
Moreover, the trajectory obtained with the proposed optimal offline scheme is more sophisticated than that of the proposed online schemes due to the non-causal knowledge of the CSI.
In addition, we also observe that the trajectories of the proposed optimal and suboptimal online schemes are identical during the entire considered time period.
On the other hand, for baseline scheme 1, the UAV always hovers above the origin as no optimization of the $(x,y)$ coordinates is performed.

\begin{figure}[t]
\centering\vspace*{-5mm}
\begin{minipage}[b]{0.45\linewidth} \hspace*{-1cm}
    \includegraphics[width=3.55in]{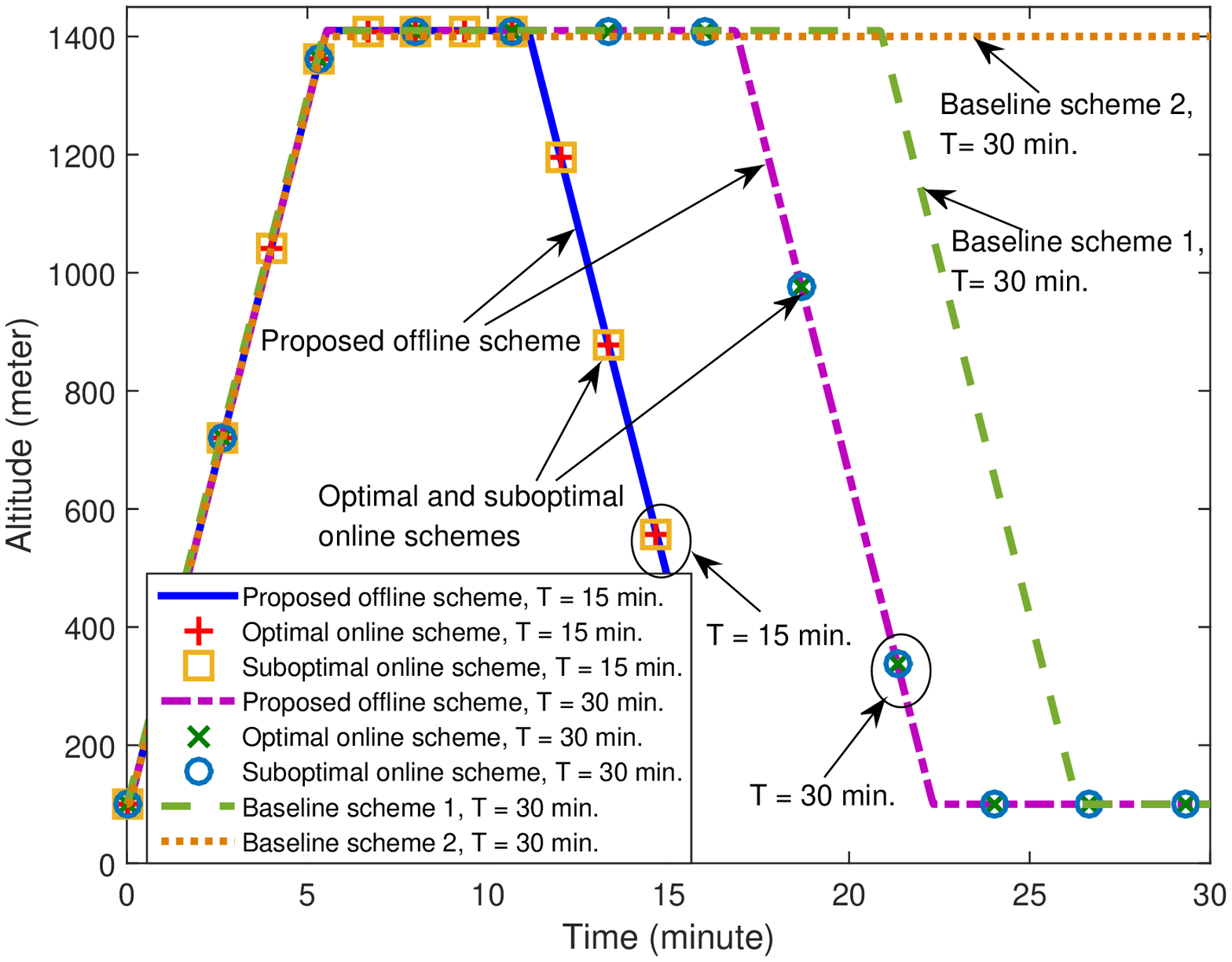}\vspace*{-7mm}
\caption{Trajectory in the vertical plane for $K = 4$ and different schemes.}
\label{fig:altitude_vs_time}
\end{minipage}\hspace*{8mm}
\begin{minipage}[b]{0.45\linewidth} \hspace*{-1cm}
    \includegraphics[width=3.55in]{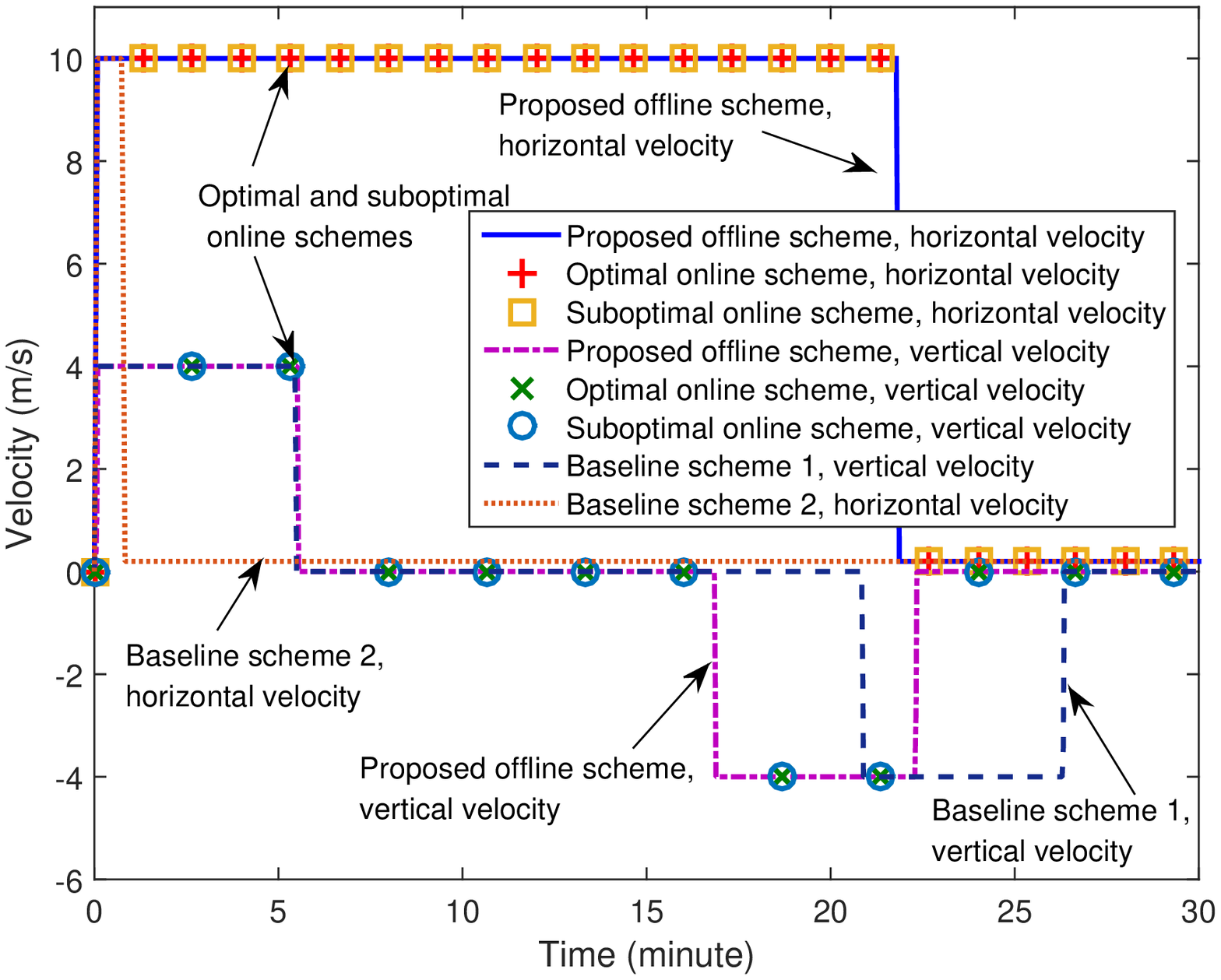}\vspace*{-7mm}
\caption{Flight velocity for $T = 30$ minutes and different schemes.}
\label{fig:velocity_vs_time}
\end{minipage}\vspace*{-4mm}
\end{figure}

Figure \ref{fig:altitude_vs_time} shows the trajectory in the vertical plane for different values of $T$ and $K=4$ users.
As can be observed, the proposed offline and online schemes yield very similar trajectories. In particular, for the proposed offline and online schemes, the UAV first climbs up until it is right above the cloud. Then, it cruises above the cloud for a certain period of time.
Subsequently, the UAV flies to a lower altitude to come closer to the users to strike a balance between the amount of energy that can be harvested and the achieved throughput.
In fact, since the UAV is energy limited, it first harvests and stores sufficient energy in the battery above the cloud and then descends to a lower altitude which leads to a smaller path loss for the communication links between the UAV and the users and thereby improves the system sum throughput.
%%%
Besides, for longer periods $T$, the offline and online schemes force the UAV to stay above the clouds for a longer time since more energy is required to sustain the operation of the UAV for longer flight periods.
%%%%
In addition, from Figure \ref{fig:altitude_vs_time}, we also observe that the vertical trajectory of the proposed suboptimal online algorithm closely approaches that of the proposed optimal online algorithm.
%%%%%
On the other hand, for baseline scheme 1, the UAV hovers above the cloud for a longer time duration compared to the proposed offline and online schemes. In fact, since for baseline scheme 1, the UAV is fixed in the horizontal plane, i.e., $\norm{\big(v_x[n], v_y[n]\big)} = 0$, a higher aerodynamic power is consumed compared to the proposed offline and online schemes. Thus, for baseline scheme 1, the UAV needs to harvest more solar energy.
For baseline scheme 2, as expected, the UAV climbs to altitude $z=\min\{L_{\mathrm{up}},z_{\mathrm{min}}\}$ and stays at the fixed altitude for the rest of the considered period.

\vspace*{-2mm}
\subsection{Velocity}

Figure \ref{fig:velocity_vs_time} shows the horizontal and the vertical velocities of the UAV during the considered period of $T =30$ minutes for different resource allocation schemes and $K=4$.
As can be observed, for the vertical velocity, for the proposed offline and online schemes, the UAV flies at the maximum vertical speed $\abs{V_{\mathrm{max}}^{\mathrm{z}}}$ whenever climbing up or down.
On the other hand, for the horizontal velocity, for the proposed offline and online schemes, the UAV flies with the maximum horizontal velocity $V_{\mathrm{max}}^{\mathrm{xy}}$ when it flies above the clouds in order to reduce the aerodynamic power consumption.
However, the UAV reduces the speed to $0.1$ m/s when flying at the lowest altitude since an excessively large horizontal velocity may lead to a large horizontal displacement which causes large fluctuations of the path loss of the communication links and a degradation of the system throughput.
%%%%%%
Moreover, we also observe that the proposed optimal and suboptimal online schemes yield a similar velocity control policy as the proposed offline scheme.
%%%%
In addition, baseline scheme 1 obtains a vertical velocity policy during the climbing phase similar to that of the proposed schemes. Yet, it prefers a longer time for energy harvesting before flying down since it consumes more aerodynamic power compared to the other schemes.
For baseline scheme 2, the UAV speeds up until it reaches a favorable horizontal position for maximizing the system sum throughput and then hovers there during the rest of the period.

\begin{figure}[t]
\centering\vspace*{-5mm}
\begin{minipage}[b]{0.45\linewidth} \hspace*{-1cm}
    \includegraphics[width=3.55in]{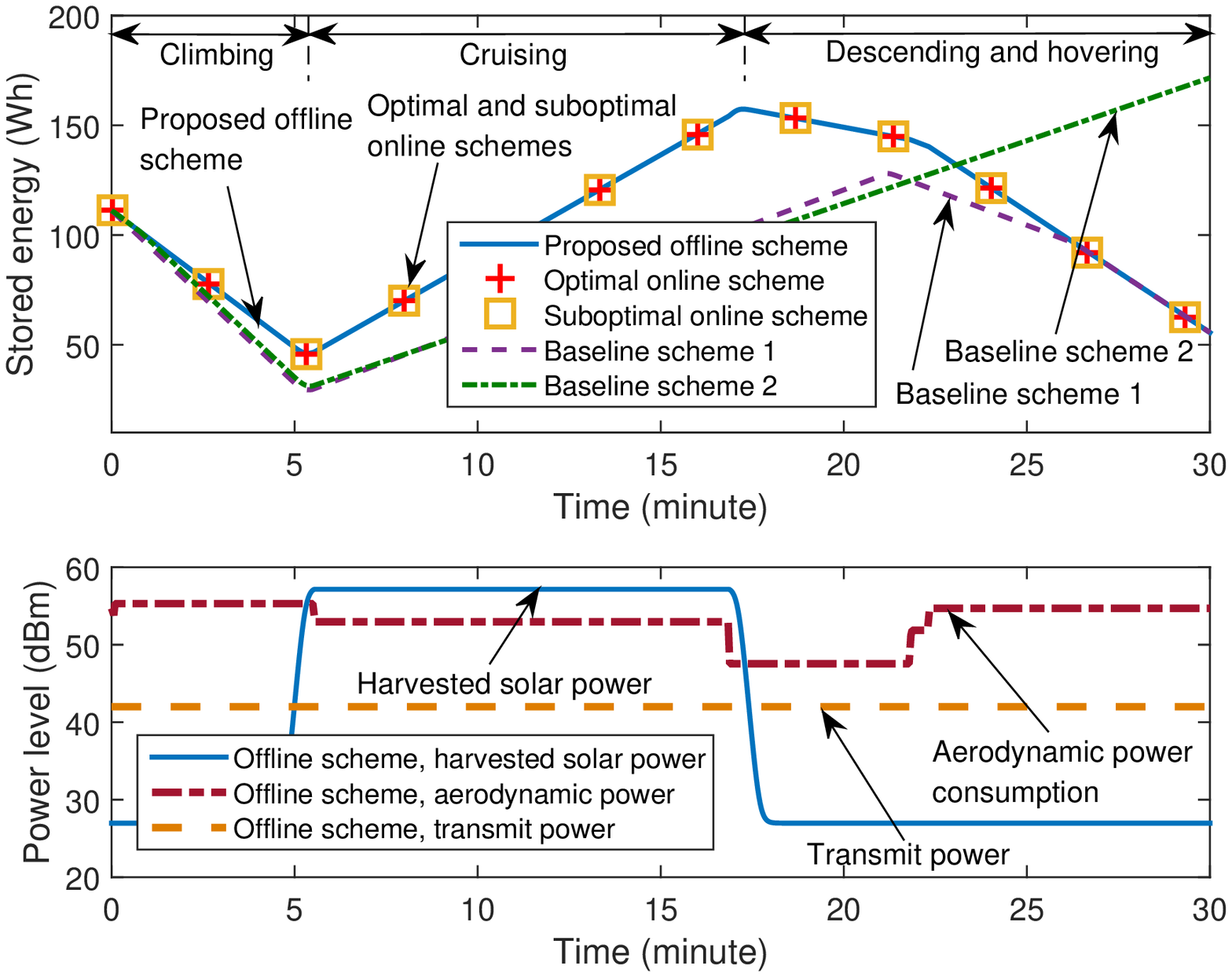}\vspace*{-7mm}
\caption{Stored energy (top figure) for different resource allocation schemes and the consumed and harvested power (bottom figure) for the proposed optimal offline scheme.}
\label{fig:energy_vs_time}
\end{minipage}\hspace*{8mm}
\begin{minipage}[b]{0.45\linewidth} \hspace*{-1cm}
    \includegraphics[width=3.55in]{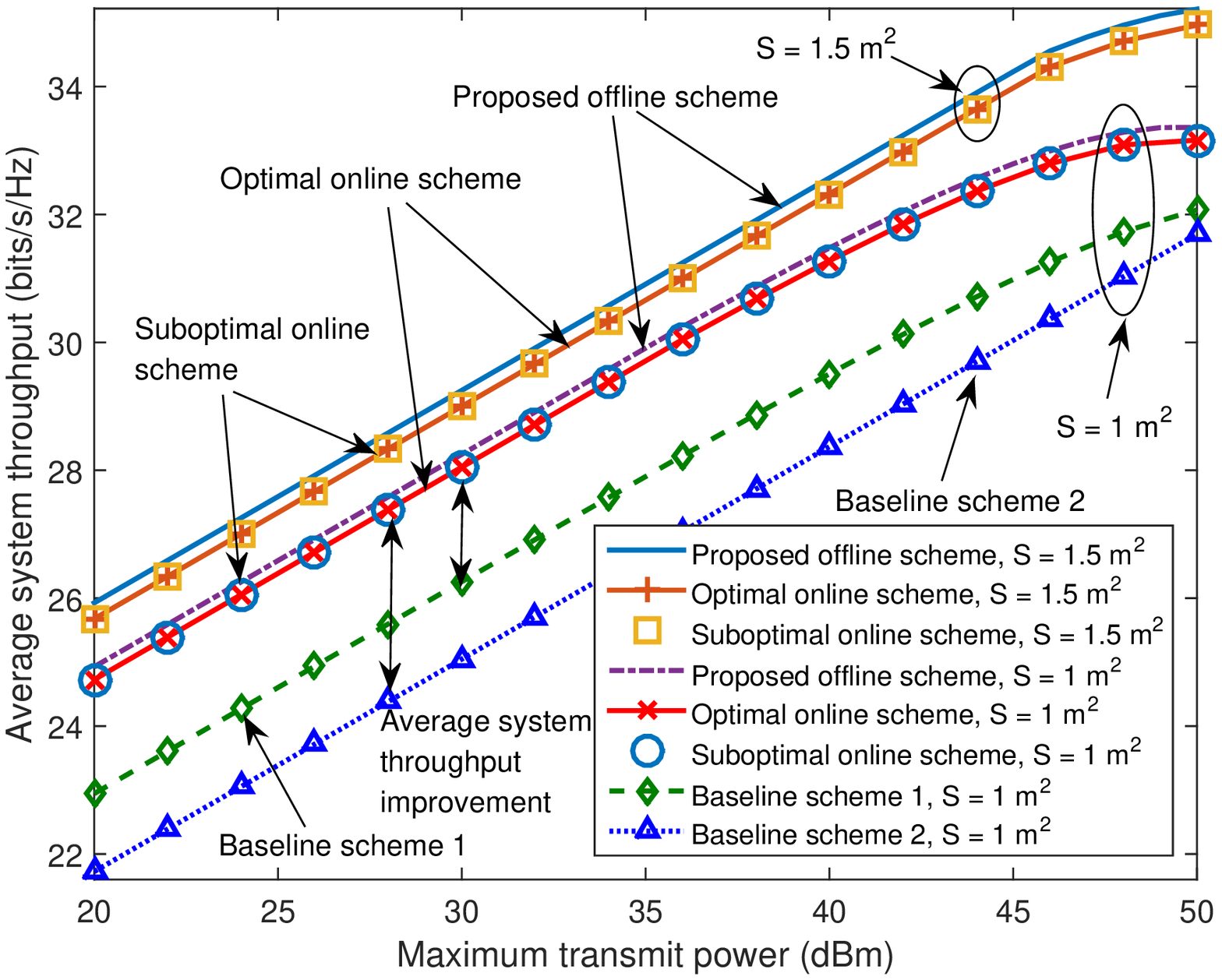} \vspace*{-14mm}
\caption{Average system throughput (bits/s/Hz) versus maximum transmit power of the UAV (dBm), $P_{\mathrm{max}}$, for different schemes with $K=4$, $q_{0} = 167 $ $\mathrm{Wh}$, and $q_{\mathrm{end}} = 278 $ $\mathrm{Wh}$. }
\label{fig:throughput_vs_power}
\end{minipage}\vspace*{-8mm}
\end{figure}

\vspace*{-2mm}
\subsection{Stored Energy, Consumed Power, and Harvested Power}
In Figure \ref{fig:energy_vs_time}, for the proposed offline scheme, we study the stored energy $\mathbf{q}$ during the considered period of $T = 30$ minutes for different resource allocation schemes and show the consumed power and the harvest power $\underline{P}^{\mathrm{solar}}\big(z[n]\big)$.
%%%%%
As can be seen, the stored energy of all considered schemes decreases until the UAV reaches a sufficiently high altitude where energy harvesting is more efficient. Then, the energy in the battery increases as the UAV flies above the clouds where the harvested solar power is higher than the total power consumption. The stored energy decreases again until it reaches the required $q_{\mathrm{end}}$ when the UAV descents and the period available for transmission expires.
We notice that the energies stored for the proposed optimal and suboptimal online schemes are similar to that for the proposed offline scheme. In other words, they are expected to also have similar resource allocation policies.
Besides, the stored energy of baseline scheme 1 is lower than that of the proposed schemes since more aerodynamic power is consumed due to the fixed horizontal position.
%Baseline scheme 2 stores more power than the proposed scheme at the end of the period since it always flies above the clouds and thus can store considerably more amount of solar energy.
Although baseline scheme 2 has more stored energy than the proposed schemes at $n=N_{\mathrm{T}}$ as it operates always above the clouds, its performance is worse than that of the proposed schemes in terms of the system sum throughput due to an exceedingly large path loss.
%%%%%%%%
Furthermore, we also show the aerodynamic power consumption, the transmit power consumption, and the harvested power of the proposed offline scheme in Figure \ref{fig:energy_vs_time} (bottom figure).
As can be observed, for the adopted system parameters, for the proposed offline scheme, the UAV always transmits signals with the maximum transmit power $P_{\mathrm{max}}$ over the entire period $T$ in order to maximize the system sum throughput.
Besides, the aerodynamic power consumption is determined by the flight status.  It can be observed that the aerodynamic power consumption of the proposed offline scheme follows the velocity changes shown in Figure \ref{fig:velocity_vs_time}.
On the other hand, Figure \ref{fig:energy_vs_time} shows that the UAV can harvest a considerable amount of solar energy when it is flying right above the clouds, cf. Figure \ref{fig:altitude_vs_time}.

\vspace*{-2mm}
\subsection{Average System Throughput versus Transmit Power}
In Figure \ref{fig:throughput_vs_power}, we investigate the average system throughput versus the maximum transmit power of the UAV, $P_{\mathrm{max}}$, for $K=4$ users, $q_{0} = 167 $ $\mathrm{Wh}$, $q_{\mathrm{end}} = 278 $ $\mathrm{Wh}$, and different solar panel sizes $S$.
In particular, the average system throughput is calculated as $\frac{{\sum}_{n = 1}^{N_{\mathrm{T}}}  {\sum}_{i = 1}^{N_{\mathrm{F}}}  {\sum}_{k = 1}^{K}  R_k^i[n]}{N_{\mathrm{T}}\mathcal{W}}$. As can be observed, the average system throughputs of the proposed offline and online schemes increase monotonically with the maximum transmit power $P_{\mathrm{max}}$ since they can effectively exploit the increased transmit power allowance to improve the received signal-to-noise ratio (SNR) at the users.
Besides, there is a diminishing return in the average system throughput if $P_{\mathrm{max}}$ exceeds $45$ dBm. In fact, for high transmit power consumptions, the UAV has to collect and store more energy in the battery, leading to longer hovering times above the clouds. Hence, for a given period of $T$, there is less time left for the UAV to operate at a low altitude to serve the users under a smaller path loss for air-to-ground communications, which partially neutralizes the improvement in the average system throughput introduced by a higher $P_{\mathrm{max}}$.
%%%%%%%%
In addition, for a larger value of $S$, the proposed offline and online schemes achieve a higher average system throughput. In fact, since the output power of the solar panels is directly proportional to the size of the solar panel, a UAV equipped with a larger solar panel needs less time to harvest the same amount of solar energy as a smaller solar panel.
As a result, the UAV can descent earlier and transmit for a longer time over low-path loss channels to the users.
Moreover, as expected, the proposed offline scheme achieves a higher average system throughput than the proposed online schemes.
%%%%%%%%
Furthermore, as can be observed, the proposed schemes achieve considerably higher average system throughputs than baseline schemes $1$ and $2$ due to the joint optimization of the 3D trajectory and the power and subcarrier allocation.
In particular, for baseline scheme $1$, the horizontal coordinates of the UAV are fixed, leading to a higher aerodynamic power consumption and less energy for data communication which hampers the average system throughput. For baseline scheme $2$, although the adopted random subcarrier allocation policy and the fixed flying altitude provide resource allocation fairness and guarantee a sufficient power supply, respectively, they result in a poor utilization of the system resources and severe path loss for air-to-ground communication, respectively.

\vspace*{-2mm}
\subsection{Average System Throughput versus Number of Users}
\begin{figure}[t]
\centering\vspace*{-5mm}
\begin{minipage}[b]{0.45\linewidth} \hspace*{-10mm}
 \includegraphics[width=3.55in]{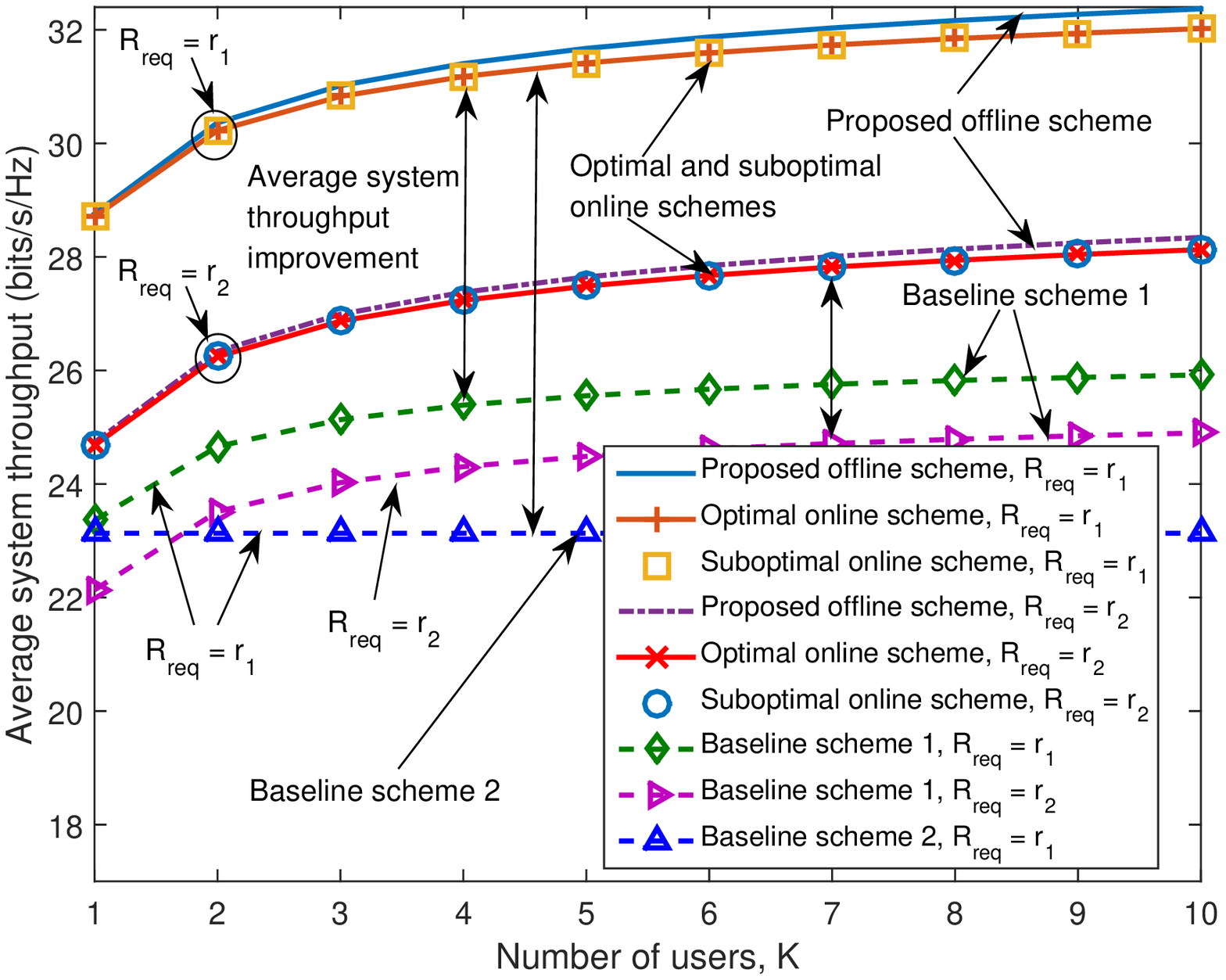} \vspace*{-12mm}
\caption{Average system throughput (bits/s/Hz) versus number of users for different resource allocation schemes, $q_{0} = 167 $ $\mathrm{Wh}$, and $q_{\mathrm{end}} = 278$ $\mathrm{Wh}$. }
\label{fig:throughput_vs_users}
\end{minipage}\hspace*{8mm}
\begin{minipage}[b]{0.45\linewidth} \hspace*{-10mm}
   \includegraphics[width=3.55in]{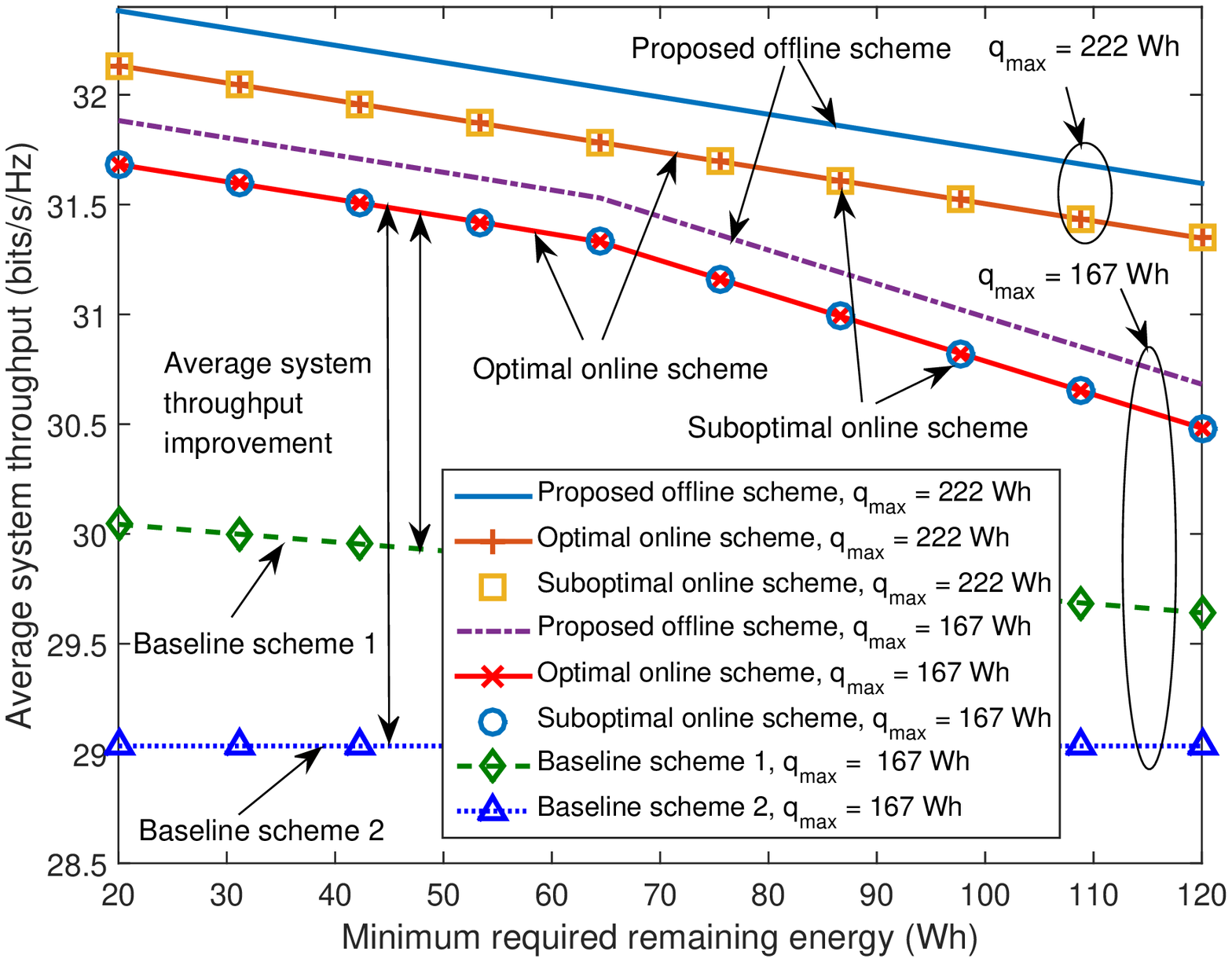} \vspace*{-12mm}
\caption{Average system sum throughput (bits/s/Hz) versus minimum required remaining energy for different resource allocation schemes and $P_{\mathrm{max}}\hspace*{-0.5mm}=\hspace*{-0.5mm}42$ dBm. }
\label{fig:throughput_vs_q_end}
\end{minipage}\vspace*{-4mm}
\end{figure}

In Figure \ref{fig:throughput_vs_users}, we investigate the average system throughput versus the total number of users $K$ for $q_{0} = 167$ $\mathrm{Wh}$, $q_{\mathrm{end}} = 278$ $\mathrm{Wh}$, and different values of $R_{\mathrm{req}}$, i.e., $r_1=50$ Mbits/s and $r_2=150$ Mbits/s.
The results shown in this section are averaged over different realizations of path loss and multipath fading where the users are randomly distributed within the cell.
%%%%%
As can be observed, the average system throughput for the proposed offline and online schemes and baseline scheme $1$ increase with the number of users since these schemes are able to exploit multiuser diversity.
However, the performance of baseline scheme $2$ is independent of the number of users since it employs a random subcarrier allocation policy.
Besides, as can be observed from Figure \ref{fig:throughput_vs_users}, the average system throughput of the proposed schemes grows faster with the number of users than that of baseline scheme $1$.
In fact, for baseline scheme $1$, the UAV cannot adjust its horizontal coordinates $(x,y)$ which limits its capability to exploit multiuser diversity by moving towards users to improve the channel conditions.
%%%
In addition, the proposed offline scheme achieves a higher average system throughput than the optimal online scheme due to the non-causal knowledge of the channel gains.
%%%
Moreover, the performance of the proposed suboptimal online scheme closely approaches that of the proposed optimal online scheme, even for a relatively large value of $K$.
%%%
Furthermore, the proposed schemes and baseline scheme 1 achieve a lower average system throughput when the QoS requirements become more stringent. In fact, to satisfy the higher minimum date rate requirements, the maximum cruising altitude is reduced to alleviate the propagation path loss to the users. As a consequence, the UAV may harvest a smaller amount of solar energy and thereby has to prolong the harvesting duration at high altitude, leaving less time for low attitude communication which reduces the average system throughput.

\vspace*{-2mm}
\subsection{Average System Throughput versus Minimum Required Remaining Energy}
In Figure \ref{fig:throughput_vs_q_end}, we investigate the average system throughput versus the minimum required remaining energy $q_{\mathrm{end}}$, for different values of the maximum storage capacity $q_{\mathrm{max}}$ and different resource allocation schemes.
As can be seen, the average system throughput of all considered schemes (except baseline scheme 2) decrease monotonically with $q_{\mathrm{end}}$.
This is because the UAV has to collect more solar energy over the entire period to achieve the larger required remaining energy. As a result, the UAV is forced to fly at a high altitude for a longer duration which degrades the average system throughput.
On the other hand, for a smaller maximum storage capacity $q_{\mathrm{max}}$, the proposed schemes also achieve a lower average system throughput.
In fact, for smaller values of $q_{\mathrm{max}}$, the UAV can store less energy in the battery.
Once the battery is fully charged, the UAV flies to a lower altitude for a certain period of time and then has to climb up again as the stored energy is not sufficient for the rest of the period.
%%%%%%%%%
%Thus, the UAV has to stay at the high altitude for a longer time until they are close enough to the users and the stored energy is sufficient for supporting the power consumption of the remaining operating.
%%%%%%%
In addition, we observe that the proposed suboptimal online scheme achieves a similar performance as the optimal online scheme. Also, as expected, the proposed offline scheme outperforms the proposed online schemes due to the availability of non-causal knowledge of the channel gains.
%%%%%
Furthermore, the performance of baseline scheme 2 is insensitive to $q_{\mathrm{end}}$ due to the fixed altitude setting. In particular, for the considered altitude, it can harvest sufficient solar energy to ensure that the remaining stored energy is larger than the minimum required value, $q_{\mathrm{end}}$.
%%%%

\vspace*{-1mm}
\section{Conclusions}
In this paper, we investigated the jointly optimal 3D trajectory, power adaptation, and subcarrier allocation algorithm design for solar-powered MC-UAV communication systems.
Due to the propagation properties of solar light and wireless signals, there is a fundamental tradeoff between harvesting solar energy, trajectory energy consumption, and communication performance.
To study this tradeoff, we first focused on the optimal offline resource allocation design by assuming non-causal knowledge of the channel gains. The objective of the formulated mixed-integer non-convex optimization problem was to maximize the system sum throughput over a finite operation period taking into account the aerodynamic power consumption, the solar energy harvesting, the finite on-board energy storage, and the minimum QoS requirements of the users. Exploiting tools from monotonic optimization theory, the offline resource allocation problem was solved optimally. Then, we studied the online resource allocation design which only requires causal CSI. The optimal online resource allocation algorithm design was developed to unveil the optimal system performance. Also, an iterative suboptimal online scheme was proposed to strike a balance between computational complexity and optimality. Simulation results revealed that the performance of the proposed suboptimal online scheme closely approaches that of the offline scheme. Moreover, our results show that to maximize the system sum throughput, the solar-powered UAV first climbs up to a high altitude to harvest a sufficient amount of solar energy before it descents to a lower altitude to shorten the communication distance to the users. Finally, the proposed offline and online resource allocation schemes achieve a significant improvement in system performance compared to the two considered baseline schemes.

\vspace*{-0mm}
\section*{Appendix-Proof of Theorem 1}
Without loss of generality, we focus on the power allocation of subcarrier $i$ in time slot $n$. For convenience, we drop index $[n]$ from the optimization variables to simplify the notation. Assume the total transmit power allocated to subcarrier $i$ is $\overline{P}^i$ and $\sum_{j=1}^K \tilde{p}_{j}^i = \overline{P}^i$. We also define $a_k^i \triangleq \sum_{j\neq k}^K \tilde{p}_{j}^i$. The achievable data rate of user $k$ on subcarrier $i$ in \eqref{rate-k-eqv} can be rewritten as \vspace*{-1mm}
\begin{eqnarray}
R_k^i&=&  \log_2 \Big( 1 + \frac{H_k^i \tilde{p}_k^i}{\xi H_k^i a_k^i + \theta_k} \Big) = \log_2 \Big( 1 + \frac{ \overline{P}^i - a_k^i}{\xi a_k^i + \frac{\theta_k}{H_k^i}} \Big).
\end{eqnarray}
For $a_k^i \neq 0$ and a sufficiently large constant value $\xi$, e.g. $\xi \gg 1$ and $\xi a_k^i \to \infty$,  we have  \vspace*{-1mm}
\begin{eqnarray}\label{a_neq0}
R_k^i=\log_2 \Big( 1 + \frac{ \overline{P}^i - a_k^i}{\xi a_k^i + \frac{\theta_k}{H_k^i}} \Big) = 0.
\end{eqnarray}
Besides, for $a_k^i = 0$  and non-zero $\overline{P}^i$, we have \vspace*{-1mm}
\begin{eqnarray}\label{a_eq0}
R_k^i=\log_2 \Big( 1 + \frac{H_k^i \overline{P}^i}{\theta_k} \Big) > 0.
\end{eqnarray}
If there are $N \ge 2$ users multiplexed on subcarrier $i$, according to \eqref{a_neq0}, we have \vspace*{-1mm}
\begin{eqnarray}\label{sumR_a_neq0}
a_k^i \neq 0, \forall k\in\{1,\ldots,K\} \quad \Longrightarrow \quad R_k^i = 0, \forall k \quad \Longrightarrow \quad \sum_{m=1}^K R_m^i =0.
\end{eqnarray}
If only user $k$ is assigned to subcarrier $i$, by combining \eqref{a_neq0} and \eqref{a_eq0}, we have \vspace*{-1mm}
\begin{eqnarray}\label{sumR_a_eq0}
a_k^i = 0,a_j^i \neq 0, \forall j\neq k \quad \Longrightarrow \quad R_k^i > 0, R_j^i = 0, \forall j\neq k \quad \Longrightarrow \quad \sum_{m=1}^K R_m^i =R_k^i > 0.
\end{eqnarray}
Therefore, the optimal solution of the system sum throughput maximization problem in \eqref{equiv-prob} will assign at most one user to each subcarrier. \hfill\qed

\vspace*{-0mm}
\bibliographystyle{IEEEtran}
\bibliography{UAV_Path_plann_solar_MC}

\end{document}